\documentclass[aps,prd,12point,twocolumn,nofootinbib,showpacs,superscriptaddress]{revtex4-2}

\usepackage{graphicx}
\usepackage{dcolumn}
\usepackage{tabu}
\usepackage{amsmath,amssymb}
\usepackage{float}
\usepackage{natbib}
\usepackage{balance}
\usepackage[normalem]{ulem}
\usepackage{braket}
\usepackage{upgreek}
\usepackage{mathrsfs}
\usepackage{bm}
\usepackage{lipsum}
\usepackage{comment}
\usepackage{xfrac}
\usepackage[dvipsnames]{xcolor}
\usepackage[title]{appendix}
\usepackage{txfonts}
\usepackage{xcolor}
\usepackage{slashed}
\usepackage[breaklinks=true,pdfborder={0 0 0},bookmarksopen]{hyperref}
\hypersetup{
	colorlinks   = true, 
	urlcolor     = magenta, 
	linkcolor    = blue, 
	citecolor   = Red 
}

\begin{document}

\title{Curved-spacetime dynamics of spin-$\frac{1}{2}$ particles in superposed states from a WKB approximation of the Dirac equation}

\author{F. Hammad}
\email{fhammad@ubishops.ca}
\affiliation{Department of Physics \& Astronomy,
Bishop's University,\\
2600 College Street, Sherbrooke, QC, J1M 1Z7, Canada}
\affiliation{Physics Department, Champlain College-Lennoxville,\\
2580 College Street, Sherbrooke, QC, J1M 0C8, Canada}

\author{M. Simard}
\email{msimard23@ubishops.ca}
\affiliation{Department of Physics \& Astronomy,
Bishop's University,\\
2600 College Street, Sherbrooke, QC, J1M 1Z7, Canada}

\author{R. Saadati}
\email{rsaadati@ubishops.ca}
\affiliation{Department of Physics \& Astronomy,
Bishop's University,\\
2600 College Street, Sherbrooke, QC, J1M 1Z7, Canada}

\author{A. Landry}
\email{a.landry@dal.ca}
\affiliation{Department of Mathematics and Statistics, Dalhousie University, Halifax, NS B3H\,4R2, Canada}

\begin{abstract}
We investigate the dynamics of spin-$\frac{1}{2}$ particles that are freely propagating in superposed states in curved spacetime. We first make use of a Wentzel–Kramers–Brillouin approximation of the Dirac equation in curved spacetime to extract the corresponding Mathisson-Papapetrou-Dixon equations that describe the deviation from geodesic motion as well as the spin precession of such particles. We then discuss, in light of our results, the case of flavour neutrinos which are, by nature, a superposition of mass eigenstates.
\end{abstract}

\maketitle
\section{Introduction}\label{sec:Intro}
The behavior of extended classical bodies under the influence of gravity within the framework of general relativity has been extensively studied in the literature very early on. That study culminated in the formulation of two sets of equations, derived mainly by Mathisson \cite{Mathisson} and Papapetrou \cite{Papapetrou}, and put on more general and solid foundations by Dixon \cite{Dixon1964,Dixon1970a,Dixon1970b} (see also Ref.\,\cite{Dixon2008} for a nice account of all the intermediate contributions from various other authors.) The first set of equations describes the deviation from geodesic motion of classical spinning bodies, whereas the second set of equations describes the spin dynamics of those bodies as they move inside a gravitational field. The Mathisson-Papapetrou-Dixon (MPD) equations {\textemdash}\,as they have come to be known\, {\textemdash} emerge from the equation of conservation of energy and momentum of the spinning body combined with a multipole expansion of the body's energy-momentum tensor.

Given that quantum particles possess an intrinsic spin angular momentum, we naturally expect such particles to also exhibit, whenever they propagate inside a gravitational field, a deviation from geodesic motion as well as a spin dynamics just as dictated by the MPD equations. However, although the MPD equations {\textemdash}\,as extended later by Dixon \cite{Dixon1964}\,{\textemdash} allow one to incorporate a contribution from intrinsic spin angular momentum, the latter has to be added into the equations by hand. One simply inserts the canonical spin tensor one obtains from the extended body's field transformation under spacetime coordinate transformations. No quantum mechanical \textit{derivation} of the intrinsic spin contribution is ever provided within the framework of those equations themselves. Therefore, directly applying the MPD equations to study quantum particles' motion just because intrinsic spin can be field-theoretically incorporated in a classical way into the equations is not fully satisfactory. Indeed, given that the dynamics of spin-$\frac{1}{2}$ particles is fully governed by the Dirac equation, the effect of gravity on the motion of such particles cannot be assumed to be the same as the one that would emerge from the \textit{classical} MPD equations even when the latter are amended by a term coming from the canonical spin tensor.

The task of deriving MPD-like equations for spin-$\frac{1}{2}$ particles by starting from the Dirac equation was taken up by R\"udiger \cite{Rudiger} and Audretsch \cite{Audretsch} who extracted their equations by applying a Wentzel–Kramers–Brillouin (WKB) approximation method to the Dirac equation in curved spacetime\footnote{For an approach based on a Foldy–Wouthuysen transformation of the Dirac Hamiltonian, see Refs.\,\cite{SilenkoTeryaev,ObukhovSilenkoTeryaev}. For an approach based on the eikonal approximation combined with a Gaussian wavepacket, see Refs.\,\cite{CianfraniMontani1,CianfraniMontani2}}. The application of a WKB approximation to the Dirac equation in Minkowski spacetime was initiated by Pauli \cite{Pauli}, and improved later by Rubinow and Keller \cite{RubinowKeller} and Rafanelli and Schiller \cite{RafanelliSchiller}, before it got generalised to curved spacetime for particles coupled to the Maxwell field as well as for massless particles (see Ref.\,\cite{Oancea} and references therein.) Moreover, besides making the extracted equations automatically anchored to more solid quantum foundations (by being extracted from the Dirac equation), the WKB approach offers an invaluable tool for taking into account the other important feature of quantum particles, which is the ability of the latter to be in a superposition of different quantum states.

The classical MPD equations have actually been extensively put to use in the literature \cite{Dvornikov2006,Alavi1,Alavi2,PRD2021,NSO} for studying the spin precession of neutrinos which, as is well known, are not only quantum particles endowed with intrinsic spin, but are also made of a superposition of different quantum states that allow them to undergo flavor oscillations as well (see the more recent works \cite{Buoninfante,Sadjadi,NSOPacket,Swami,NSOPW,Dvornikov2023} and the references therein.) In fact, the three flavor neutrino states detected experimentally so far are made of a superposition of three different mass eigenstates, each of which carries a different inertial mass. This observation puts therefore into full perspective our discussion above. Indeed, applying the WKB approximation to the Dirac equation for extracting MPD-like equations for particles that propagate as a superposition of different quantum states becomes of paramount importance for studying any physical phenomenon that involves the dynamics of such particles in curved spacetime. In this paper, we have set ourselves the goal of carrying out such a task.      

We organized the remainder of this paper as follows. In Sec.\,\ref{sec:MPDIntro}, we briefly introduce the classical MPD equations and recall what each of the different terms displayed in those equations means. In Sec.\,\ref{sec:MPDfromWKB}, we give a review of the derivation of the MPD-like equations for spin-$\frac{1}{2}$ particles from a WKB approximation of the Dirac equation in curved spacetime based on Refs.\,\cite{Rudiger,Audretsch}. We give a derivation of those equations that neither appeals to a symplectic Hamiltonian (see Ref.\,\cite{Oancea} and references therein) nor requires any specific choice of frame \cite{Rudiger,Audretsch}. Our derivation of those equations will indeed be tailored to easily accommodate the multi-state scenario that is of interest to us here. In Sec.\,\ref{sec:SuperposedMPD}, we use the results and the tools of Sec.\,\ref{sec:MPDfromWKB} to derive the dynamics of a spin-$\frac{1}{2}$ particle that freely propagates as a superposition of two different quantum states in curved spacetime. A rigorous elaboration on the subtleties coming from defining a dynamical 4-momentum to be associated with such particles is provided. We then make use of the results of Sec.\,\ref{sec:SuperposedMPD} to discuss in Sec.\,\ref{sec:Application} the case of flavor neutrinos that are made of a superposition of different mass eigenstates. We summarise and discuss our main findings in a brief conclusion given in Sec.\,\ref{sec:Conclusion}. More detailed steps of some of the calculations required in the text are collected in appendices \ref{App1} to  \ref{App8}.
\section{MPD equations for classical spinning bodies}\label{sec:MPDIntro}
For a later comparison between the classical and quantum dynamics of particles with spin, we devote this section to displaying the classical MPD equations and briefly recalling the definition of the various terms they contain. This will also allow us to fix some of the notation to be used throughout the rest of this paper. 

Let $p^\mu$ be the dynamical 4-momentum of a spinning body, that should not be confused with the body's kinematical 4-momentum $\pi^\mu=m\varv^\mu$ that satisfies\footnote{We use, throughout the paper, the spacetime metric signature $(-,+,+,+)$ and we set $c=1$.} $\pi_\mu\pi^\mu=-m^2$. Here, $m$ is the body's mass, and $\varv^\mu={\rm d}x^\mu/{\rm d}\tau$ is the body's center-of-mass 4-velocity ($\varv_\mu \varv^\mu=-1$) for any affine parameter $\tau$ that is taken to be the proper time of the body. Then, at the pole-dipole approximation (which consists of considering only the momentum and the spin angular momentum of the body, and ignoring the higher multipole moments of the latter), the first set of MPD equations describes the non-geodesic motion of the spinning body, and reads
\begin{equation}\label{MPD1}
\dot{p}^\mu\equiv\frac{{\rm D}p^\mu}{{\rm d}\tau}=-\tfrac{1}{2}R^\mu_{\;\;\nu\rho\sigma}\varv^\nu S^{\rho\sigma}.
\end{equation}
Here, we introduced the dot notation (which we shall use throughout the paper) to denote the proper time derivative. The operator ${\rm D}={\rm d}x^\mu\nabla_\mu$ stands for the total covariant derivative, $R^\mu_{\;\;\nu\rho\sigma}$ is the Riemann curvature tensor, and $S^{\mu\nu}$ is the spin tensor in which is encoded the spin-angular momentum vector $S^\mu$ of the body according to $S^\mu=-\frac{1}{2m}\epsilon^\mu_{\;\;\nu\rho\lambda}p^\nu S^{\rho\lambda}$. The totally antisymmetric Levi-Civita tensor $\epsilon_{\mu\nu\rho\lambda}=\sqrt{-g}\,\varepsilon_{\mu\nu\rho\lambda}$ is given in terms of the metric determinant $g$ and the Levi-Civita alternating symbol $\varepsilon_{\mu\nu\rho\lambda}$, normalised  such that $\varepsilon_{0123}=1$. 

The second set of MPD equations describes the dynamics of the spin tensor $S^{\mu\nu}$, and reads
\begin{equation}\label{MPD2}
\dot{S}^{\mu\nu}\equiv\frac{{\rm D}S^{\mu\nu}}{{\rm d}\tau}=p^\mu \varv^\nu-p^\nu \varv^\mu.
\end{equation}
Note that the right-hand side of this equation does not vanish as the 4-vectors $p^\mu$ and $\varv^\mu$ are not proportional to each other. Note also that contracting both sides of Eq.\,(\ref{MPD1}) with $\varv_\mu$ yields $\varv_\mu\dot{p}^\mu=0$, by means of which one deduces the conservation of mass $\dot{m}=0$ after performing the identification $m=-\varv_\mu p^\mu$. Therefore, contracting both sides of Eq.\,(\ref{MPD2}) with $\varv_\nu$ yields $p^\mu=m\varv^\mu-\varv_\nu \dot{S}^{\mu\nu}$, which provides at the first order in spin the explicit relation between the dynamical 4-momentum $p^\mu$ and the kinematical 4-momentum $m\varv^\mu$. Furthermore, at the zeroth order in spin, the first set of equations (\ref{MPD1}) reduces to the geodesic equation $\dot{p}^\mu=0$ of freely propagating spinless point particles, whereas the second set of equations (\ref{MPD2}) reduces to $\dot{S}^{\mu\nu}=0$, which describes the well-known gyroscope precession in free fall \cite{MTW}.

Besides equations (\ref{MPD1}) and (\ref{MPD2}), the so-called Tulczyjew-M{\o}ller condition $S^{\mu\nu}p_\nu=0$ \cite{Moller,Tulczyjew} (as opposed to the so-called Pirani-Mathisson condition $S^{\mu\nu}\varv_\nu=0$ \cite{Mathisson,Pirani}) is also imposed on the classical spin tensor $S^{\mu\nu}$ through the dynamical 4-momentum $p^{\mu}$ of the body. This condition is imposed as a \textit{supplementary} condition to supply us with the extra three equations required to make the system of seven independent equations (\ref{MPD1}) and (\ref{MPD2}) determinate enough to solve for the ten unknowns $\varv^\mu$, $\varv_\mu p^\mu$ and $S^{\mu\nu}$. For a further discussion on these and other supplementary conditions, see Ref.\,\cite{KyrianSemerak} and references therein. 
\section{MPD-like equations from a WKB approximation}\label{sec:MPDfromWKB}
The purpose of this section is to give a review of the method for extracting MPD-like equations from the curved-spacetime Dirac equation that describes the dynamics of a single spinor field in curved spacetime. The derivations that will be given here follow closely the works of R\"udiger and Audretsch \cite{Rudiger,Audretsch}. The slight difference, though, is that we shall deal here with labeled states, we shall not rely on any specific reference frame, and we shall extract a few extra equations that are general enough to provide us with the 
tools 
without which the case of superposed spinor fields of Sec.\,\ref{sec:SuperposedMPD} cannot be tackled within our approach. Furthermore, in view of our application of the results of this section to the case of flavor neutrinos, we consider here the case of spinor fields that carry different masses.

We use in what follows Latin subscripts $(i,j)$ to distinguish the different possible states of a particle, each carrying a different mass. Then, the Dirac equation for a spin-$\frac{1}{2}$ particle in a state described by the spinor field $\Psi_i(x)$, carrying a mass $m_i$ and freely propagating in curved spacetime, reads
\begin{equation}\label{DiracEq}
\left(i\hbar\gamma^\mu\nabla_\mu-m_i\right)\Psi_i(x)=0.
\end{equation}
Note that throughout this paper no summation is intended when repeated Latin indices $i$ and $j$ are displayed.  The curved-spacetime gamma
matrices $\gamma^\mu$ are built from the flat-spacetime constant gamma matrices $\gamma^a$ by projecting the latter onto the curved manifold using the vierbeins $e^\mu_a$. The latter are defined with the help of the Minkowski metric $\eta^{ab}$ by $e^\mu_ae^\nu_b\eta^{ab}=g^{\mu\nu}$ for any spacetime metric $g_{\mu\nu}$ of inverse $g^{\mu\nu}$. The spin-covariant derivative acting on spinor fields is defined by $\nabla_\mu=\partial_\mu+\tfrac{1}{8}\omega^{ab}_\mu[\gamma_a,\gamma_b]$.
The spin connection $\omega_\mu^{ab}$ is built from the vierbeins and the Christoffel symbols extracted from the metric as $\omega_\mu^{\,\,ab}=-e^{\nu b}\partial_\mu e^a_\nu+e^{\nu b}\Gamma_{\mu\nu}^\lambda e^{a}_\lambda$.

Plugging the WKB ansatz for the spinor field $\Psi_i(x)$ \cite{Rudiger,Audretsch},
\begin{equation}\label{WKBAnsatz}
    \Psi_i(x)=\exp\left[\frac{i}{\hbar}\mathcal{S}_i(x)\right]\sum_{n=0}^\infty\hbar^n\psi_i^{(n)}(x),
\end{equation}
where $\mathcal{S}_i(x)$ is a real scalar function and $\psi_i^{(n)}(x)$ are four-component spinors, into the Dirac equation (\ref{DiracEq}), and then equating to zero the coefficient of each power of $\hbar$, one easily extracts the following set of equations for the spinors $\psi^{(n)}_i(x)$ \cite{Rudiger,Audretsch}:
\begin{align}
    &\left(\gamma^\mu\pi_{i\mu}+m_i\right)\psi_i^{(0)}=0\label{0thOrder},\\
&\left(\gamma^\mu\pi_{i\mu}+m_i\right)\psi_i^{(n)}=i\gamma^\mu\nabla_\mu\psi_i^{(n-1)},\quad n=1,2,3,\ldots\label{1stOrder}
\end{align}
We set $\pi_{i\mu}=\partial_\mu \mathcal{S}_i$, to be identified with the kinematical 4-momentum of the particle in the state described by the spinor field $\Psi_i(x)$. Eq.\,(\ref{0thOrder}) is an algebraic equation that has a nontrivial solution $\psi_i^{(0)}$ when $\det(\gamma^\mu\pi_{i\mu}+m_i)=0$. This condition yields the mass-shell equation $\pi_{i\mu} \pi_i^\mu=-m_i^2$ that gives the usual geodesic equation of a particle of mass $m_i$ and of 4-momentum $\pi_i^\mu$. This equation is also the Hamilton-Jacobi equation $\partial_\mu \mathcal{S}_i\partial^\mu \mathcal{S}_i=-m_i^2$ for the phase function $\mathcal{S}_i(x)$. Therefore, by setting $
\pi_i^\mu=m_i\varv^\mu$, we conclude that, to the zeroth order in $\hbar$, the particle simply follows a geodesic of tangent 4-vector $\varv^\mu$. In other words, spin has no effect on the trajectory of the particle at the zeroth order in $\hbar$. 

As the matrix multiplying $\psi_i^{(0)}$ in Eq.\,(\ref{0thOrder}) is a rank-2 matrix, the two linearly independent eigenspinors $\Theta_{A}(x)$ and $\Theta_{B}(x)$ of such a matrix imply that a general eigenspinor $\psi_i^{(0)}$ is a linear combination of the form \cite{Rudiger,Audretsch},
\begin{equation}\label{IndependentSolutions}
\psi_i^{(0)}(x)=a_i^{(0)}(x)\Theta_{A}(x)+b_i^{(0)}(x)\Theta_{B}(x),
\end{equation}
for some complex scalar factors $a_i^{(0)}(x)$ and $b_i^{(0)}(x)$. Note that the 4-spinors $\Theta_A(x)$ and $\Theta_B(x)$ do not carry an index $i$ because the mass $m_i$ in Eq.\,(\ref{0thOrder}) can be factored out and the matrix multiplying these 4-spinors simply reads $\gamma^\mu \varv_\mu+\mathbb{I}$, where $\mathbb{I}$ denotes here (and henceforth) the 4$\times$4 identity matrix. Also, being linearly independent and normalized, the two 4-spinors $\Theta_{A}(x)$ and $\Theta_{B}(x)$ satisfy the following identities \cite{Rudiger,Audretsch}:
\begin{equation}\label{ThetaA&B}
\bar{\Theta}_{A}\Theta_{B}=\delta_{AB},\qquad
\bar{\Theta}_{A}\gamma^\mu\Theta_{B}=\varv^\mu\delta_{AB}.    
\end{equation}
The second of these two identities follows from the first one after plugging the combination (\ref{IndependentSolutions}) back into Eq.\,(\ref{0thOrder}). In addition, we find that these 4-spinors also obey the following two constraints:
\begin{equation}\label{DotThetaA&DotThetaB}
\dot{\Theta}_{A}=C_{1}\Theta_{A}+C_{2}\Theta_{B},\qquad\dot{\Theta}_{B}=D_{1}\Theta_{B}+D_{2}\Theta_{A},
\end{equation}
where $C_{1}$, $C_{2}$, $D_{1}$ and $D_{2}$ are four arbitrary complex scalars. The constraints (\ref{DotThetaA&DotThetaB}) on these four complex scalars, as well as the nature of the latter, are all derived in detail in Appendix \ref{App1}. Moreover, when combining the two constraints (\ref{DotThetaA&DotThetaB}) with Eqs.\,(\ref{0thOrder}) and (\ref{ThetaA&B}), we extract the following additional identities to be satisfied by the 4-spinors (see Appendix \ref{App2} for a detailed derivation):
\begin{align}\label{BisThetaA&B}
\bar{\Theta}_{A}\gamma^\mu\nabla_\mu\Theta_{A}&=\tfrac{1}{2}\nabla_\mu \varv^\mu+C_{1},\quad 
\bar{\Theta}_{B}\gamma^\mu\nabla_\mu\Theta_{B}=\tfrac{1}{2}\nabla_\mu \varv^\mu+D_{1},\nonumber\\
\bar{\Theta}_{A}\gamma^\mu\nabla_\mu\Theta_{B}&=D_{2}.    
\end{align}

On the other hand, the 4-spinors $\Theta_{A}(x)$ and $\Theta_{B}(x)$ still need to be constrained by the solvability conditions of the remaining equation (\ref{1stOrder}). Indeed, Eq.\,(\ref{1stOrder}) is a nonhomogeneous linear equation that is solvable if and only if the linearly independent solutions $\bar{\Theta}_{A}(x)$ and $\bar{\Theta}_{B}(x)$ of the transposed homogeneous version of the equation are both orthogonal to the term causing the non-homogeneity of the equation. In other words, we need to have the two conditions
$\bar{\Theta}_{A}\gamma^\mu\nabla_\mu\psi_i^{(n-1)}=0$ and $\bar{\Theta}_{B}\gamma^\mu\nabla_\mu\psi_i^{(n-1)}=0$ satisfied for any integer $n\geq1$ as well.
Upon inserting the linear combination (\ref{IndependentSolutions}) into these extra two conditions after setting $n=1$, the latter translate into:
\begin{align}
\dot{a}_i^{(0)}&=-\left(\tfrac{1}{2}\nabla_\mu \varv^\mu+C_{1}\right)a_i^{(0)}-D_{2}\,b_i^{(0)},\nonumber\\
\dot{b}_i^{(0)}&=-\left(\tfrac{1}{2}\nabla_\mu \varv^\mu+D_{1}\right)b_i^{(0)}-C_{2}\,a_i^{(0)}.\label{Nabla(a)Nabla(b)}
\end{align}
Upon using Eqs.\,(\ref{DotThetaA&DotThetaB}) and (\ref{Nabla(a)Nabla(b)}), we learn that each of the zeroth-order spinor fields $\psi_i^{(0)}(x)$ obeys the following dynamics \cite{Rudiger,Audretsch}:
\begin{equation}\label{Dotpsi(o)}
\dot{\psi}_i^{(0)}(x)=-\tfrac{1}{2}\left(\nabla_\mu \varv^\mu\right)\psi_i^{(0)}(x).
\end{equation}

Similarly, being solutions to the nonhomogeneous equations (\ref{1stOrder}) the 4-spinors $\psi_i^{(n)}(x)$ (for $n\geq1$) can be written as linear combinations of the 4-spinors $\Theta_{A}(x)$, $\Theta_{B}(x)$ and $\xi_i^{(n)}(x)$. The spinors $\xi_i^{(n)}(x)$ are orthogonal to $\Theta_{A}(x)$ and $\Theta_{B}(x)$ and are the particular solutions of Eq.\,(\ref{1stOrder}). Thus, for $n\geq1$ we have \cite{Rudiger},
\begin{align}\label{psi(n)}
\psi_i^{(n)}(x)&=a_i^{(n)}(x)\Theta_{A}(x)+b_i^{(n)}(x)\Theta_{B}(x)+\xi_i^{(n)}(x),\nonumber\\
n&=1,2,3,\ldots    
\end{align}
where $a_i^{(n)}(x)$ and $b_i^{(n)}(x)$ are arbitrary complex scalar factors to be constrained. The explicit expressions of the 4-spinors $\xi_i^{(n)}(x)$ can be found in terms of two other mutually orthogonal 4-spinors $\Pi_{A}(x)$ and $\Pi_{B}(x)$ that are also orthogonal to the 4-spinors $\Theta_{A}(x)$ and $\Theta_{B}(x)$. Therefore, $\Theta_A(x)$, $\Theta_B(x)$, $\Pi_A(x)$ and $\Pi_B(x)$ form a complete orthonormal 4-spinor basis in terms of which our solutions $\psi_i^{(n)}(x)$ can be expressed. For that purpose, we can easily check that it is sufficient to have $\Pi_{A}(x)$ and $\Pi_{B}(x)$ be the two linearly independent solutions of the equation $\left(\gamma^\mu \varv_{\mu}-\mathbb{I}\right)\Pi=0$. It follows then that these two spinors also satisfy the following two identities \cite{Rudiger}:
\begin{equation}\label{PiA&B}
\bar{\Pi}_{A}\Pi_{B}=-\delta_{AB},\qquad
\bar{\Pi}_{A}\gamma^\mu\Pi_{B}=\varv^\mu\delta_{AB}.    
\end{equation}
By plugging the combination (\ref{psi(n)}) into Eq.\,(\ref{1stOrder}) and making use of the identities (\ref{PiA&B}), we extract the following explicit expression of the 4-spinors $\xi_i^{(n)}(x)$ \cite{Rudiger}:
\begin{align}\label{XiSpinor}
\xi_i^{(n)}&=-\frac{i}{2m_i}\left\{\left[\bar{\Pi}_{A}\gamma^\mu\nabla_\mu\psi_i^{(n-1)}\right]\Pi _{A}+\left[\bar{\Pi}_{B}\gamma^\mu\nabla_\mu\psi_i^{(n-1)}\right]\Pi_{B}\right\},\nonumber\\
n&=1,2,3,\ldots
\end{align}

In analogy to the constraints (\ref{DotThetaA&DotThetaB}) we got on the 4-spinors $\Theta_{A}(x)$ and $\Theta_{B}(x)$, we also derive by means of the same method the following constraints on the 4-spinors $\Pi_{A}(x)$ and $\Pi_{B}(x)$:
\begin{equation}\label{DotPiA&DotPiB}
\dot{\Pi}_{A}=K_{1}\Pi_{A}+K_{2}\Pi_{B},\qquad
\dot{\Pi}_{B}=L_{1}\Pi_{B}+L_{2}\Pi_{A},
\end{equation}
where $K_{1}$, $K_{2}$, $L_{1}$ and $L_{2}$ are also four arbitrary complex scalars that are of the same nature as the scalars $C_{1}$, $C_{2}$, $D_{1}$ and $D_{2}$; namely, $K_{1}=-K^*_{1}$, $L_{1}=-L^*_{1}$ and $K_{2}=-L^*_{2}$. From the two equations (\ref{DotPiA&DotPiB}), we extract the following two additional identities satisfied by these two 4-spinors:
\begin{align}\label{BisPiA&B}
\bar{\Pi}_{A}\gamma^\mu\nabla_\mu\Pi_{A}&=\tfrac{1}{2}\nabla_\mu \varv^\mu+K_{1},\quad 
\bar{\Pi}_{B}\gamma^\mu\nabla_\mu\Pi_{B}=\tfrac{1}{2}\nabla_\mu \varv^\mu+L_{1},\nonumber\\
\bar{\Pi}_{A}\gamma^\mu\nabla_\mu\Pi_{B}&=L_{2}.    
\end{align}

The complex scalar factors $a_i^{(n)}(x)$ and $b_i^{(n)}(x)$ entering the combination (\ref{psi(n)}) are constrained by the conditions $\bar{\Theta}_{A}\gamma^\mu\nabla_\mu\psi_i^{(n)}=0$ and $\bar{\Theta}_{B}\gamma^\mu\nabla_\mu\psi_i^{(n)}=0$ for $n\geq1$. The latter two conditions together with the combination (\ref{psi(n)}) yield,
\begin{align}
\dot{a}_i^{(n)}&=-\left(\tfrac{1}{2}\nabla_\mu \varv^\mu+C_{1}\right)a_i^{(n)}-D_{2}b_i^{(n)}-\bar{\Theta}_{A}\gamma^{\mu}\nabla_\mu\xi_i^{(n)},\nonumber\\
\dot{b}_i^{(n)}&=-\left(\tfrac{1}{2}\nabla_\mu \varv^\mu+D_{1}\right)b_i^{(n)}-C_{2}a_i^{(n)}-\bar{\Theta}_{B}\gamma^{\mu}\nabla_\mu\xi_i^{(n)}.\label{Dot(a)&Dot(b)}
\end{align}
Combining the constraints (\ref{DotPiA&DotPiB}) and (\ref{Dot(a)&Dot(b)}), we deduce the following dynamical equation for the $n$th-order spinor field $\psi_{i}^{(n)}(x)$ when $n\geq1$:
\begin{align}\label{Dotpsi(n)}
\dot{\psi}_i^{(n)}&=-\tfrac{1}{2}\left(\nabla_\mu \varv^\mu\right)\psi_i^{(n)}+\tfrac{1}{2}\left(\nabla_\mu \varv^\mu\right)\xi^{(n)}_i\nonumber\\
&\quad-\left(\bar{\Theta}_{A}\gamma^{\mu}\nabla_\mu\xi_i^{(n)}\right)\Theta_{A}-\left(\bar{\Theta}_{B}\gamma^{\mu}\nabla_\mu\xi_i^{(n)}\right)\Theta_{B}+\dot{\xi}^{(n)}_i.
\end{align}
From this general equation, the following two identities, that are necessary for deriving the main formulas of Sec.\,\ref{sec:SuperposedMPD}, are easily derived by setting $n=1$ (see Appendix \ref{App3} for the detailed steps):
\begin{equation}\label{psi0Dotpsi1Andpsi1Dotpsi0}
\bar{\psi}_i^{(0)}\dot{\psi}_j^{(1)}=-\tfrac{1}{2}(\nabla_\mu \varv^\mu)\bar{\psi}_i^{(0)}\psi_j^{(1)},\quad \bar{\psi}_i^{(1)}\dot{\psi}_j^{(0)}=-\tfrac{1}{2}(\nabla_\mu \varv^\mu)\bar{\psi}_i^{(1)}\psi_j^{(0)}.
\end{equation}

Consider now the Gordon decomposition of the conserved Dirac 4-current $\bar{\Psi}_i\gamma^\mu\Psi_i$ \cite{Rudiger,Audretsch}:
\begin{equation}\label{GordonDecomp}
    j_i^\mu(x)=\frac{i\hbar}{2m_i}\left(\nabla^\mu\bar{\Psi}_i\Psi_i-\bar{\Psi}_i\nabla^\mu\Psi_i\right)+\frac{\hbar}{2m_i}\nabla_\nu\left(\bar{\Psi}_i\sigma^{\mu\nu}\Psi_i\right),
\end{equation}
where we used the customary notation for the commutator of the gamma matrices: $\sigma^{\mu\nu}\equiv\frac{i}{2}[\gamma^\mu,\gamma^\nu]$. The first term in this sum is identified with the convection 4-current $j_{ic}^\mu(x)$ associated with the state $\Psi_i(x)$, whereas the second term is identified with the corresponding spin 4-current $j^\mu_{is}(x)$. Both currents are separately conserved thanks to the Dirac equation (\ref{DiracEq}). From the convection 4-current, one extracts the dynamical 4-momentum $p_i^\mu=m_ij_{ic}^{\mu}/(\bar{\Psi}_i\Psi_i)$ which, after inserting into it the WKB ansatz (\ref{WKBAnsatz}), takes, up to the first order in $\hbar$, the following form \cite{Rudiger,Audretsch}:
\begin{equation}\label{QuantumDynamical4-momentum}
 p_i^\mu(x)=\pi_i^\mu+\frac{i\hbar}{2\bar{\psi}_i^{(0)}\psi_i^{(0)}}\left[\nabla^\mu\bar{\psi}_i^{(0)}\psi_i^{(0)}-\bar{\psi}_i^{(0)}\nabla^\mu\psi_i^{(0)}\right].
\end{equation}
Thus, the various identities derived above, together with the geodesic equation $\dot{\pi}_i^\mu=0$ satisfied by the kinematical 4-momentum, we easily show, as worked out in detail in Appendix \ref{App4}, that
\begin{equation}\label{QMPD1}
\dot{p}_i^\mu=-\tfrac{1}{2}R^\mu_{\;\;\nu\rho\sigma}\varv^\nu S_i^{\rho\sigma}-(\nabla^\mu \varv_{\nu}) p_i^\nu.  
\end{equation}
Here, $S_i^{\mu\nu}$ is the spin tensor associated to the particle in the state described by the field $\Psi_i(x)$. The explicit definition of that tensor in terms of $\Psi_i(x)$ reads \cite{Rudiger,Audretsch}, 
\begin{equation}\label{SpinTensor}
S_i^{\mu\nu}=\hbar\frac{\bar{\Psi}_i\sigma^{\mu\nu}\Psi_i}{2\bar{\Psi}_i{\Psi}_i}=\hbar\frac{\bar{\psi}_i^{(0)}\sigma^{\mu\nu}\psi_i^{(0)}}{2\bar{\psi}_i^{(0)}{\psi}_i^{(0)}}+\mathcal{O}(\hbar^2).
\end{equation}
In the second step, we have kept only the leading term that is first-order in $\hbar$. We may take, as is done in Ref.\,\cite{Audretsch} for a single state, the components $S^{\mu\nu}_i$ of the tensor (\ref{SpinTensor}) to represent the components of the spin per particle in the state $\Psi_i(x)$, for in the non-relativistic limit $\Psi_i(x)$ is a spinor wavefunction and one may interpret $\bar{\Psi}_i\sigma^{\mu\nu}\Psi_i$ and $\bar{\Psi}_i\Psi_i$ as the spin density and the number of particles density in that state, respectively. Using the definition (\ref{SpinTensor}), we straightforwardly compute the proper time derivative $\dot{S}_i^{\mu\nu}$ by making use of Eq.\,(\ref{Dotpsi(o)}). To first order in $\hbar$, the result is \cite{Rudiger,Audretsch}:
\begin{equation}\label{QMPD2}
    \dot{S}_i^{\mu\nu}=0.
\end{equation}

The two results (\ref{QMPD1}) and (\ref{QMPD2}) are the quantum analogs of the first set and second set of the classical MPD equations (\ref{MPD1}) and (\ref{MPD2}), respectively. One immediately notices the difference between the two sets of equations. Concerning the first set of equations, we note that although the term on the right-hand side of the classical equations (\ref{MPD1}) is fully recovered in Eq.\,(\ref{QMPD1}), the extra term $-(\nabla^\mu \varv_{\nu}) p_i^\nu$ that is not present on the right-hand side of Eq.\,(\ref{MPD1}) also arises here. In Refs.\,\cite{Rudiger,Audretsch} the left-hand side of Eq.\,(\ref{QMPD1}) is rather written as $m^{-1}p^\nu\nabla_\nu p^\mu$, which, to first order in $\hbar$, precisely amounts to having $\dot{p}^\mu$ plus the extra term $p^\nu\nabla_\nu \varv^\mu$ which leads, thanks to Eq.\,(\ref{Dotpsi(o)}), to our second term on the right-hand side of Eq.\,(\ref{QMPD1}). It is also important for our discussion in the next section to notice here that at the zeroth order in $\hbar$, Eq.\,(\ref{QMPD1}) reduces, as expected, to the geodesic equation $\dot{p}_i^\mu=0$ of a classical point particle. 

Concerning the second set of equations, we also note that the right-hand side of the classical equations (\ref{MPD2}) does not vanish, in contrast to the right-hand side of the quantum-mechanical result (\ref{QMPD2}), even though we were careful in keeping in the derivation of the latter all terms that are up to the first order in $\hbar$. Note that these two slight discrepancies between the two sets of classical and quantum equations do not arise within a purely Lagrangian approach \cite{Pavsic,Barut,GaioliGarcia,NuriUnal}. Finally, we easily check that the definition (\ref{SpinTensor}) of the spin tensor yields, at first order in $\hbar$, the Tulczyjew-M{\o}ller condition $ S^{\mu\nu}_ip_{i\nu}=0$ for each state $\Psi_i(x)$ thanks to Eq.\,(\ref{ThetaA&B}). 
\section{MPD-like equations for superposed states}\label{sec:SuperposedMPD}
Consider a particle of mass $m_I$ freely propagating in curved spacetime as a single spinor field $\Phi_I(x)$ made of a linear superposition of two different spinor fields $\Psi_1(x)$ and $\Psi_2(x)$ carrying masses $m_1$ and $m_2$, respectively. The effective mass $m_I$ of the particle is a function of the masses $m_1$ and $m_2$ carried by the two fields $\Psi_1(x)$ and $\Psi_2(x)$.
We assume the following decomposition of the spinor field $\Phi_I(x)$ in terms of the fields $\Psi_1(x)$ and $\Psi_2(x)$: 
\begin{align}\label{PhiI&PhiII}
    \Phi_I(x)&=\cos\theta\,\Psi_1(x)+
\sin\theta\,\Psi_2(x),\nonumber\\
\Phi_{II}(x)&=-\sin\theta\,\Psi_1(x)+
\cos\theta\,\Psi_2(x).
\end{align}
The real parameter $\theta$ of the superposition is taken to be a constant in spacetime. The second superposition $\Phi_{II}(x)$, obtained from the first via an orthogonal rotation in the state space $(\Psi_1,\Psi_2)$, is automatically associated to a second particle of mass $m_{II}$. Being in a superposition of two different states, the dynamical 4-momentum of each of the two particles contains interference terms arising from the overlap of the two fields $\Psi_1(x)$ and $\Psi_2(x)$. There are now three different options for extracting a 4-momentum that could play the role played by the dynamical 4-momentum of Eq.\,(\ref{QMPD1}).

\subsection{First option for a dynamical 4-momentum}
One option for building the 4-momenta is to assign $p_I^{A\mu}(x)$ and $p_{II}^{A\mu}(x)$ to the first and the second particle, respectively, by using convection 4-currents that would be associated to the superpositions $\Phi_I(x)$ and $\Phi_{II}(x)$ simply by substituting in the first term on the right-hand side of formula (\ref{GordonDecomp}) the spinor field $\Psi_i(x)$ by the fields $\Phi_I(x)$ and $\Phi_{II}(x)$, respectively. We use here a superscript $A$ to denote the momenta obtained within this first option in order to distinguish them from the momenta we obtain within the second and third options dealt with below. The 4-momentum, say $p_{I}^{A\mu}(x)$, would then read
\begin{equation}\label{SuperposedDynamicalP}
p_I^{A\mu}=\frac{i\hbar}{2\bar{\Phi}_I\Phi_I}\left(\nabla^\mu\bar{\Phi}_I\Phi_I-\bar{\Phi}_I\nabla^\mu\Phi_I\right).
\end{equation}
The 4-momentum $p_{II}^{A\mu}$ would be obtained from this expression by the subscript substitution $I\rightarrow II$. On the other hand, from the definition (\ref{SuperposedDynamicalP}) of the 4-momentum $p_I^{A\mu}$, we  can express the latter in terms of the momenta $p_1^{\mu}$ and $p_2^\mu$ as
\begin{align}\label{pIAinp1p2}
p_I^{A\mu}&=\frac{\bar{\Psi}_1\Psi_1}{\bar{\Phi}_I\Phi_I}p_1^\mu\cos^2\theta+\frac{\bar{\Psi}_2\Psi_2}{\bar{\Phi}_I\Phi_I}p_2^\mu\sin^2\theta\nonumber\\
&+\frac{i\hbar\sin2\theta}{4\bar{\Phi}_I\Phi_I}\left(\nabla^\mu\bar{\Psi}_1\Psi_2-\bar{\Psi}_1\nabla^\mu\Psi_2+\nabla^\mu\bar{\Psi}_2\Psi_1-\bar{\Psi}_2\nabla^\mu\Psi_1\right).
\end{align}

Using our results from the previous section concerning the dynamics of the individual 4-spinors $\Psi_i(x)$, we find that the dynamical 4-momentum $p_I^{A\mu}$ of the particle obeys in this case the following equation of motion: 
\begin{widetext}
\begin{equation}\label{Dotp}
\dot{p}_I^{A\mu}=-\frac{1}{2} R^{\mu}_{\;\;\nu\rho\lambda}\varv^\nu S_I^{\rho\lambda}-(\nabla^\mu \varv_{\nu})p_I^{A\nu}+\left(\frac{i}{\hbar}\frac{\bar{\Psi}_1\Psi_2-\bar{\Psi}_2\Psi_1}{2\bar{\Phi}_I\Phi_I}\, p_I^{A\mu}+\frac{\nabla^\mu\bar{\Psi}_1\Psi_2-\bar{\Psi}_1\nabla^\mu\Psi_2+\bar{\Psi}_2\nabla^\mu\Psi_1-\nabla^\mu\bar{\Psi}_2\Psi_1}{4\bar{\Phi}_I\Phi_I}\right)_{\mathcal{O}(\hbar)}\Delta m_{21}\sin2\theta+\mathcal{O}(\hbar^2).
\end{equation}
\end{widetext}
The subscript $\mathcal{O}(\hbar)$ on brackets means here (and henceforth) that we keep inside the brackets only those terms that are at most of the indicated order in $\hbar$. Furthermore, we introduced, for convenience, the notation $\Delta m_{21}=m_2-m_1$. The detailed steps leading to this result are given in Appendix \ref{App5}. Here, we defined, in analogy to Eq.\,(\ref{SpinTensor}), the effective spin tensor $S_I^{\mu\nu}$ associated to the particle in the quantum superposition $\Phi_I(x)$ by
\begin{equation}\label{SpinS}
S_I^{\mu\nu}=\frac{\hbar\bar{\Phi}_I\sigma^{\mu\nu}\Phi_I}{2\bar{\Phi}_I{\Phi}_I}.
\end{equation}
In terms of the spin tensors $S_1^{\mu\nu}$ and $S_2^{\mu\nu}$ of the single-state particles, this spin tensor reads
\begin{align}
S_I^{\mu\nu}&=\frac{\bar{\Psi}_1\Psi_1}{\bar{\Phi}_I\Phi_I}S_1^{\mu\nu}\cos^2\theta+\frac{\bar{\Psi}_2\Psi_2}{\bar{\Phi}_{I}\Phi_{I}}S_2^{\mu\nu}\sin^2\theta\nonumber\\
&\quad+\frac{\hbar}{4\bar{\Phi}_I\Phi_I}(\bar{\Psi}_1\sigma^{\mu\nu}\Psi_2+\bar{\Psi}_2\sigma^{\mu\nu}\Psi_1)\sin2\theta.
\end{align}
To first order in $\hbar$, the proper time derivative of this tensor is found to be (see Appendix \ref{App6} for a detailed derivation):
\begin{multline}\label{DotS}  
\dot{S}_I^{\mu\nu}=\left(\frac{i}{\hbar}\frac{\bar{\Psi}_1\Psi_2-\bar{\Psi}_2\Psi_1}{2\bar{\Phi}_I\Phi_I} S_I^{\mu\nu}+i\frac{\bar{\Psi}_2\sigma^{\mu\nu}\Psi_1-\bar{\Psi}_1\sigma^{\mu\nu}\Psi_2}{4\bar{\Phi}_I\Phi_I}\right)_{\mathcal{O}(\hbar)}\\
\times\Delta m_{21}\sin2\theta+\mathcal{O}(\hbar^2).
\end{multline}
Note that the content inside the parentheses is made of terms of order $\hbar$ and higher as can easily be checked since at zeroth-order we have $\bar{\psi}_1^{(0)}\psi_2^{(0)}=\bar{\psi}_2^{(0)}\psi_1^{(0)}$ and $\bar{\psi}_2^{(0)}\sigma^{\mu\nu}\psi_1^{(0)}=\bar{\psi}_1^{(0)}\sigma^{\mu\nu}\psi_2^{(0)}$. Thus, the leading-order term on the right-hand side of Eq.\,(\ref{DotS}) is indeed of order $\hbar$. The equations giving $\dot{p}^\mu_{II}$, $S_{II}^{\mu\nu}$ and $\dot{S}_{II}^{\mu\nu}$ extracted from the superposition $\Phi_{II}(x)$ have the same forms as those in Eqs.\,(\ref{Dotp})-(\ref{DotS}); one has only to make in the latter equations the replacements $I\rightarrow II$, $\theta\rightarrow-\theta$ and $1\leftrightarrow 2$. Note, accordingly, that as the numerators of the ratios on the right-hand side of Eq.\,(\ref{DotS}) are both antisymmetric under the substitutions $\theta\rightarrow-\theta$ and $1\leftrightarrow 2$, at first order in $\hbar$ the time derivative of the sum $(\bar{\Phi}_I{\Phi}_IS_I^{\mu\nu}+\bar{\Phi}_{II}{\Phi}_{II}S_{II}^{\mu\nu})/(\bar{\Phi}_{I}{\Phi}_{I}+\bar{\Phi}_{II}{\Phi}_{II})$ vanishes as can be checked with the help of Eq.\,(\ref{DotPhiPhi}). This means that for superposed states, what obeys the analog of Eq.\,(\ref{QMPD2}) for single-state particles are not the individual spin tensors $S_I^{\mu\nu}$ and $S_{II}^{\mu\nu}$ associated to each superposition, but rather the weighted sum of those spin tensors.     

The results (\ref{Dotp}) and (\ref{DotS}) bring corrections to Eqs.\,(\ref{QMPD1}) and (\ref{QMPD2}) that are entirely due to the interference between the superposed states $\Psi_1(x)$ and $\Psi_2(x)$ making the particle. It is worth noting that these interference terms arise as a consequence of a direct overlap of the fields, as in the terms $\bar{\Psi}_i\Psi_j$, as well as an overlap via the intermediacy of the gravitational field, as in the terms $\bar{\Psi}_i\nabla^\mu\Psi_j$ and $\bar{\Psi}_i\sigma^{\mu\nu}\Psi_j$. Also, in both Eqs.\,(\ref{Dotp}) and (\ref{DotS}) the corrections are all proportional to the mass difference of the two quantum states and vanish only when the latter carry  identical masses. 

The result (\ref{Dotp}) clearly shows a distinct departure from Eq.\,(\ref{QMPD1}) of single-state particles. Such a difference consists of an extra term arising from the interference between the fields, and it is proportional to the mass difference $\Delta m_{21}$. As for the case of the spin tensors $S_I^{\mu\nu}$ and $S_{II}^{\mu\nu}$, however, one can build a combination of the momenta $p_I^{A\mu}$ and $p_{II}^{A\mu}$ that would satisfy the exact same equation as Eq.\,(\ref{QMPD1}). Indeed, thanks to the antisymmetry of the numerators of the extra terms inside the parentheses in Eq.\,(\ref{Dotp}) with respect to the substitutions $1\leftrightarrow2$ and $\theta\rightarrow-\theta$, we easily check with the help of Eq.\,(\ref{DotPhiPhi}) that the time derivative of the sum $(\bar{\Phi}_I\Phi_Ip_I^{A\mu}+\bar{\Phi}_{II}\Phi_{II}p_{II}^{A\mu})/(\bar{\Phi}_I\Phi_I+\bar{\Phi}_{II}\Phi_{II})$ satisfies the analog of Eq.\,(\ref{QMPD1}), with the weighted sum $(\bar{\Phi}_I{\Phi}_IS_I^{\mu\nu}+\bar{\Phi}_{II}{\Phi}_{II}S_{II}^{\mu\nu})/(\bar{\Phi}_{I}{\Phi}_{I}+\bar{\Phi}_{II}{\Phi}_{II})$ playing the role of the effective spin tensor. This means again that what obeys the analog of Eq.\,(\ref{QMPD1}) for single-state particles are not the individual 4-momenta $p_I^{A\mu}$ and $p_{II}^{A\mu}$ associated to each superposition, but rather the weighted sum of those 4-momenta.

On the other hand, using expression (\ref{pIAinp1p2}) of the 4-momentum together with the definition (\ref{SpinS}) of the spin tensor, we easily check that at first order in $\hbar$, we have $S_I^{\mu\nu}p^A_{I\nu}=0$, which is the analog of the Tulczyjew-M{\o}ller condition for superposed states based on this first option (\ref{SuperposedDynamicalP}) for the 4-momentum. Moreover, we straightforwardly check that not only does the second superposition also satisfy the condition $S_{II}^{\mu\nu}p^A_{II\nu}=0$, but that we even have $S_I^{\mu\nu}p^A_{II\nu}=0$ and $S_{II}^{\mu\nu}p^A_{I\nu}=0$ at first order in $\hbar$. This comes about thanks to Eq.\,(\ref{ThetaA&B}) and the fact that in $p_I^{A\mu}$ and $p_{II}^{A\mu}$ only the terms proportional to $\varv^\mu$ are zeroth-order in $\hbar$, which already fulfill the Pirani-Mathisson condition $S_r^{\mu\nu}\varv_\nu=0$ for $r=I,II$. Indeed, the 4-velocity $\varv^\mu$ is common to the different states and superposition of states thanks to the common proper time $\tau$ of the latter, as dictated by the equivalence principle, whereas the interference terms in those momenta are all first-order and higher in $\hbar$.
In what follows, we shall explore a different option for building the dynamical 4-momenta and how these various results get modified.  
\subsection{Second option for a dynamical 4-momentum}
The second option for building the dynamical 4-momenta of the first and the second particle, that we denote here by $p_I^{B\mu}(x)$ and $p_{II}^{B\mu}(x)$, respectively, is to first obtain a Gordon decomposition for the currents $\bar{\Phi}_{I}\gamma^\mu\Phi_{I}$ and $\bar{\Phi}_{II}\gamma^\mu\Phi_{II}$. These currents could then be used to extract the 4-momenta. In fact, the linear combinations (\ref{PhiI&PhiII}) lead to the following coupled Dirac equations satisfied by the spinors $\Phi_I(x)$ and $\Phi_{II}(x)$:
\begin{align}\label{CoupledDirac}
\left(i\hbar\gamma^\mu\nabla_\mu-m_{I}\right)\Phi_{I}(x)&=m_{I,II}\Phi_{II}(x),\nonumber\\
\left(i\hbar\gamma^\mu\nabla_\mu-m_{II}\right)\Phi_{II}(x)&=m_{I,II}\Phi_{I}(x).
\end{align}
The masses $m_{I}$ and $m_{II}$ of the particles as well as the coupling mass $m_{I,II}$, all emerging from combining the two uncoupled Dirac equations satisfied by $\Psi_1(x)$ and $\Psi_2(x)$, are given in terms of the masses $m_1$ and $m_2$ of the latter by 
\begin{align}\label{MassesI&II}
m_{I}&=m_1\cos^2\theta+m_2\sin^2\theta,\quad m_{II}=m_1\sin^2\theta+m_2\cos^2\theta,\nonumber\\
m_{I,II}&=\tfrac{1}{2}\Delta m_{21}\sin2\theta.
\end{align}
Using the coupled Dirac equations (\ref{CoupledDirac}), we easily derive the following decomposition of $\bar{\Phi}_I\gamma^\mu\Phi_I$:
\begin{widetext}
\begin{align}\label{GordonPhiIGammaPhiI}
 \bar{\Phi}_I\gamma^\mu\Phi_I&=\frac{i\hbar m_{II}}{2(m_Im_{II}-m^2_{I,II})}\left[\nabla^\mu\bar{\Phi}_I\Phi_I-\bar{\Phi}_I\nabla^\mu\Phi_I-\frac{m_{I,II}}{m_{II}}\left(\nabla^\mu\bar{\Phi}_{II}\Phi_{I}-\bar{\Phi}_{I}\nabla^\mu\Phi_{II}\right)\right]\nonumber\\
 &\quad+\frac{\hbar m_{II}}{2(m_Im_{II}-m^2_{I,II})}\left[\nabla_\nu(\bar{\Phi}_I\sigma^{\mu\nu}\Phi_I)-\frac{m_{I,II}}{m_{II}}\left(\nabla_\nu\bar{\Phi}_{II}\sigma^{\mu\nu}\Phi_{I}+\bar{\Phi}_{I}\sigma^{\mu\nu}\nabla_\nu\Phi_{II}\right)\right].
\end{align}
\end{widetext}
The decomposition of the current $\bar{\Phi}_{II}\gamma^\mu\Phi_{II}$ follows from Eq.\,(\ref{GordonPhiIGammaPhiI}) by the substitution $I\leftrightarrow II$. 
The terms in the first line on the right-hand side of Eq.\,(\ref{GordonPhiIGammaPhiI}), although not conserved, can be viewed as making a convection 4-current $J_{cI}^\mu(x)$ that could be used to extract a 4-momentum $p_I^{B\mu}(x)$ to be associated with the first particle. Setting $p_I^{B\mu}=m_IJ_{cI}^\mu/(\bar{\Phi}_I\Phi_I)$, we have
\begin{multline}\label{pIOption2}
p_I^{B\mu}=\frac{i\hbar m_I m_{II}}{m_Im_{II}-m^2_{I,II}}\left(\frac{\nabla^\mu\bar{\Phi}_I\Phi_I-\bar{\Phi}_I\nabla^\mu\Phi_I}{2\bar{\Phi}_I\Phi_I}\right.\\
\left.-\frac{m_{I,II}}{m_{II}}\frac{\nabla^\mu\bar{\Phi}_{II}\Phi_{I}-\bar{\Phi}_{I}\nabla^\mu\Phi_{II}}{2\bar{\Phi}_I\Phi_I}\right).
\end{multline}
The expression of $p_{II}^{B\mu}(x)$ is obtained from Eq.\,(\ref{pIOption2}) by making the substitution $I\leftrightarrow II$. As we did for the 4-momenta $p_I^{A\mu}$ and $p_{II}^{A\mu}$, we can express the 4-momentum $p_I^{B\mu}$ in terms of the 4-momenta $p_1^\mu$ and $p_2^\mu$ as follows:
\begin{align}\label{pIBinp1p2}
p_I^{B\mu}&=\frac{m_I m_{II}}{m_Im_{II}-m^2_{I,II}}\left[p_I^{A\mu}-\frac{m_{I,II}}{2m_{II}}\left(\frac{\bar{\Psi}_1\Psi_1p_1^{\mu}+\bar{\Psi}_2\Psi_2p_2^{\mu}}{\bar{\Phi}_I\Phi_I}\right)\sin2\theta\right.\nonumber\\
&\quad-\frac{i\hbar m_{I,II}}{2m_{II}\bar{\Phi}_I\Phi_I}(\nabla^\mu\bar{\Psi}_2\Psi_1-\bar{\Psi}_1\nabla^\mu\Psi_2)\cos^2\theta\nonumber\\
&\quad\left.+\frac{i\hbar m_{I,II}}{2m_{II}\bar{\Phi}_I\Phi_I}(\nabla^\mu\bar{\Psi}_1\Psi_2-\bar{\Psi}_2\nabla^\mu\Psi_1)\sin^2\theta\right].
\end{align}
A similar expression for $p_{II}^{B\mu}$ is obtained from Eq.\,(\ref{pIBinp1p2}) by the substitutions $1\leftrightarrow2$, $\theta\rightarrow-\theta$ and $I\leftrightarrow II$. With this 4-momentum, we check that at first order in $\hbar$ the Tulczyjew-M{\o}ller condition is also satisfied for both spin tensors, $S_I^{\mu\nu} p^B_{I\nu}=0$ and $S_{II}^{\mu\nu} p^B_{II\nu}=0$. Moreover, we also have the vanishing of the mixed products, $S_{II}^{\mu\nu} p^B_{I\nu}=0$ and $S_I^{\mu\nu} p^B_{II\nu}=0$. This comes about thanks again to the fact that all the interference terms in expression (\ref{pIBinp1p2}) are proportional to $\hbar$, while terms containing $\varv^\mu$ are zeroth-order in $\hbar$.

A lengthy, but straightforward calculation, then leads to the following proper time derivative of the 4-momentum:
\begin{widetext}
\begin{align}\label{DotPOption2}
    \dot{p}_I^{B\mu}&=-\frac{m_I m_{II}}{2(m_Im_{II}-m^2_{I,II})}R^{\mu}_{\;\;\nu\rho\lambda}\varv^\nu\left(S_I^{\rho\lambda}-\frac{\hbar m_{I,II}}{m_{II}}\frac{\bar{\Phi}_{II}\sigma^{\rho\lambda}\Phi_I+\bar{\Phi}_I\sigma^{\rho\lambda}\Phi_{II}}{4\bar{\Phi}_I\Phi_I}\right)_{\mathcal{O}(\hbar)}-(\nabla^\mu \varv_{\nu})p_I^{B\nu}\nonumber\\
    &\quad+\left(\frac{m_I m_{II}}{m_Im_{II}-m^2_{I,II}}\frac{\nabla^\mu\bar{\Psi}_1\Psi_2-\bar{\Psi}_1\nabla^\mu\Psi_2+\bar{\Psi}_2\nabla^\mu\Psi_1-\nabla^\mu\bar{\Psi}_2\Psi_1}{4\bar{\Phi}_I\Phi_I}+\frac{i}{\hbar}\frac{\bar{\Psi}_1\Psi_2-\bar{\Psi}_2\Psi_1}{2\bar{\Phi}_I\Phi_I}p_I^{B\mu}\right)_{\mathcal{O}(\hbar)}\Delta m_{21}\sin2\theta\nonumber\\
    &\quad+\frac{m_I m_{I,II}\Delta m_{21}}{m_Im_{II}-m^2_{I,II}}\left(\frac{\nabla^\mu\bar{\Psi}_2\Psi_1+\bar{\Psi}_1\nabla^\mu\Psi_2}{2\bar{\Phi}_I\Phi_I}\cos^2\theta+\frac{\nabla^\mu\bar{\Psi}_1\Psi_2+\bar{\Psi}_2\nabla^\mu\Psi_1}{2\bar{\Phi}_I\Phi_I}\sin^2\theta\right)_{\mathcal{O}(\hbar)}+\mathcal{O}(\hbar^2).
\end{align}
\end{widetext}
The detailed derivation of this equation is given in Appendix \ref{App7}. The time derivative $\dot{p}_{II}^{B\mu}$ of the second particle's 4-momentum is obtained from Eq.\,(\ref{DotPOption2}) by making the substitutions $I\leftrightarrow II$, $\theta\rightarrow-\theta$ and $1\leftrightarrow 2$. 
Given the antisymmetry of all the numerators of the ratios in the second and third lines of the result (\ref{DotPOption2}) with respect to the substitutions $\theta\rightarrow -\theta$ and $1\leftrightarrow2$, we immediately check again that, similarly to what is found for the 4-momenta $p_I^{A\mu}$ and $p_{II}^{A\mu}$, the weighted sum $(\bar{\Phi}_I\Phi_Ip_I^{B\mu}+\bar{\Phi}_{II}\Phi_{II}p_{II}^{B\mu})/(\bar{\Phi}_I\Phi_I+\bar{\Phi}_{II}\Phi_{II})$ does obey an equation analogous to Eq.\,(\ref{QMPD1}). However, besides the weighted sum $(\bar{\Phi}_I{\Phi}_IS_I^{\mu\nu}+\bar{\Phi}_{II}{\Phi}_{II}S_{II}^{\mu\nu})/(\bar{\Phi}_{I}{\Phi}_{I}+\bar{\Phi}_{II}{\Phi}_{II})$ playing again the role of the effective spin tensor, extra terms made of the mixed tensor $\bar{\Phi}_{II}\sigma^{\mu\nu}\Phi_I$ and its complex conjugate appear now coupled to the Riemann tensor as well. In what follows, we shall therefore consider a third option for building the dynamical 4-momentum of the particles. 
\subsection{Third option for a dynamical 4-momentum}\label{sec:SuperposedMDP-C}
The third option is to form, instead of two separate 4-momenta, a single dynamical 4-momentum $p_{I,II}^\mu(x)$ from both superpositions $\Phi_I(x)$ and $\Phi_{II}(x)$ by relying on a Gordon decomposition of the conserved total 4-current $J^{\mu}_{I,II}(x)=\bar{\Phi}_I\gamma^\mu\Phi_I+\bar{\Phi}_{II}\gamma^\mu\Phi_{II}$. 
One can easily check using Eq.\,(\ref{CoupledDirac}) that the 4-current $J_{I,II}^\mu(x)$ is indeed conserved. Furthermore, making use of the coupled Dirac equations (\ref{CoupledDirac}), we easily extract the following Gordon decomposition of such a current:
\begin{widetext}
\begin{align}\label{GordonPhiIPhiII}
 J^{\mu}_{I,II}(x)&=\frac{m_I m_{II}}{(m_Im_{II}-m^2_{I,II})}\left[\frac{i\hbar}{2m_I}\left(\nabla^\mu\bar{\Phi}_I\Phi_I-\bar{\Phi}_I\nabla^\mu\Phi_I\right)+\frac{i\hbar}{2m_{II}}\left(\nabla^\mu\bar{\Phi}_{II}\Phi_{II}-\bar{\Phi}_{II}\nabla^\mu\Phi_{II}\right)\right]\nonumber\\
 &\quad-\frac{i\hbar m_{I,II}}{2(m_Im_{II}-m^2_{I,II})}\left(\nabla^\mu\bar{\Phi}_{II}\Phi_{I}-\bar{\Phi}_{I}\nabla^\mu\Phi_{II}+\nabla^\mu\bar{\Phi}_{I}\Phi_{II}-\bar{\Phi}_{II}\nabla^\mu\Phi_{I}\right)\nonumber\\
 &\quad+\frac{\hbar m_I m_{II}m_{I,II}}{2(m_Im_{II}-m^2_{I,II})}\nabla_\nu\left(\frac{\bar{\Phi}_I\sigma^{\mu\nu}\Phi_I}{m_Im_{I,II}}+\frac{\bar{\Phi}_{II}\sigma^{\mu\nu}\Phi_{II}}{m_{II}m_{I,II}}-\frac{\bar{\Phi}_{II}\sigma^{\mu\nu}\Phi_I+\bar{\Phi}_I\sigma^{\mu\nu}\Phi_{II}}{m_Im_{II}}\right).
\end{align}
\end{widetext}
The first two lines on the right-hand side of this equation represent the convection 4-current $J^\mu_{cI,II}(x)$, whereas the last two lines represent the spin 4-current $J^\mu_{sI,II}(x)$. These two currents are separately conserved. The conservation equation $\nabla_\mu J^\mu_{cI,II}=0$ follows from using the coupled Dirac equations (\ref{CoupledDirac}) and the identities, $[\nabla_\mu,\nabla_\nu]\Phi=\frac{i}{4}R_{\mu\nu ab}\sigma^{ab}\Phi$ and $[\nabla_\mu,\nabla_\nu]\bar{\Phi}=-\frac{i}{4}R_{\mu\nu ab}\bar{\Phi}\sigma^{ab}$, satisfied by any 4-spinor $\Phi$ (see, e.g., Ref.\,\cite{SchroDirac}). The conservation equation $\nabla_\mu J_{sI,II}^\mu=0$ follows from the conservation of every individual term of that current as a consequence of the antisymmetry of the tensor $\sigma^{\mu\nu}$ in its two indices and the symmetry of the Christoffel symbols in their lower indices.

In a more condensed form, the coupled equations (\ref{CoupledDirac}) can be written as $(i\hbar\Gamma^\mu\nabla_\mu-{\rm M})\Sigma=0$, where the matrix $\Gamma^\mu$, the mass matrix ${\rm M}$ and the spinor $\Sigma$ are given by
\begin{equation}
\Gamma^\mu=\begin{pmatrix}
            \gamma^\mu & 0\\
            0 & \gamma^\mu
           \end{pmatrix},\quad {\rm M}=\begin{pmatrix}
               m_I & m_{I,II}\\
               m_{I,II} & m_{II}
           \end{pmatrix},\quad 
           \Sigma=\begin{pmatrix}
               \Phi_I\\
               \Phi_{II}
           \end{pmatrix}.    
    \end{equation}
The matrices $\Gamma^\mu$ and $\rm M$ are taken formally to be 2$\times$2 matrices acting on the spinor $\Sigma$ taken formally to be a 2-component spinor. As a consequence, the conserved convection and spin 4-currents $J^\mu_{cI,II}(x)$ and $J^\mu_{sI,II}(x)$ take the following condensed expressions as well:
\begin{align}\label{CondensedConvection&Spin}
J_{cI,II}^\mu(x)&=\tfrac{i}{2}\hbar\left(\nabla^\mu\bar{\Sigma}{\rm M}^{-1}\Sigma-\bar{\Sigma}{\rm M}^{-1}\nabla^\mu\Sigma\right),\nonumber\\
J_{sI,II}^\mu(x)&=\tfrac{1}{2}\hbar\nabla_\nu\left(\bar{\Sigma}\Gamma^{\mu\nu}{\rm M}^{-1}\Sigma\right).    
\end{align}
Here, we defined $\bar{\Sigma}\equiv\Sigma^{\dagger}\Gamma^0$ and we introduced the block-diagonal matrix $\Gamma^{\mu\nu}=\frac{i}{2}[\Gamma^\mu,\Gamma^\nu]$ having the matrix $\sigma^{\mu\nu}$ as its only nonvanishing elements sitting on its two diagonal entries. The first identity in Eq.\,(\ref{CondensedConvection&Spin}) then suggests the following possible expression for a dynamical 4-momentum: $p_{I,II}^\mu=M J_{cI,II}^\mu/(\bar{\Sigma}\Sigma)$, where $M$ stands for $\sqrt{\det({\rm M})}$: the square root of the determinant of the mass matrix $\rm M$ taken again formally to be a 2$\times$2 matrix acting on a 2-component spinor $\Sigma$. Therefore, we have $M=(m_Im_{II}-m_{I,II}^2)^{\frac{1}{2}}$, which leads to the following third option for the dynamical 4-momentum:
\begin{widetext}
\begin{align}\label{POption3}
p_{I,II}^\mu\!=\!\frac{m_I m_{II}}{\sqrt{m_1m_2}}\left(\frac{i\hbar}{2m_I}\frac{\nabla^\mu\bar{\Phi}_I\Phi_I-\bar{\Phi}_I\nabla^\mu\Phi_I}{\bar{\Sigma}\Sigma}+\frac{i\hbar}{2m_{II}}\frac{\nabla^\mu\bar{\Phi}_{II}\Phi_{II}\!-\!\bar{\Phi}_{II}\nabla^\mu\Phi_{II}}{\bar{\Sigma}\Sigma}\right)-\frac{i\hbar m_{I,II}}{\sqrt{m_1m_2}}\left(\frac{\nabla^\mu\bar{\Phi}_{II}\Phi_{I}-\bar{\Phi}_{I}\nabla^\mu\Phi_{II}+\nabla^\mu\bar{\Phi}_{I}\Phi_{II}-\bar{\Phi}_{II}\nabla^\mu\Phi_{I}}{2\bar{\Sigma}\Sigma}\right).
\end{align}
\end{widetext}
This expression might be cast in terms of the 4-momenta of the previous subsections as follows:
\begin{align}\label{pIIIinp1p2}
p_{I,II}^\mu&=\frac{m_Im_{II}}{\sqrt{m_1m_2}}\left(\frac{\bar{\Phi}_I\Phi_I}{m_I\bar{\Sigma}\Sigma}p_I^{A\mu}+\frac{\bar{\Phi}_{II}\Phi_{II}}{m_{II}\bar{\Sigma}\Sigma}p_{II}^{A\mu}\right)\nonumber\\
&+\frac{m_{I,II}}{\sqrt{m_1m_2}}\left(\frac{\bar{\Psi}_1\Psi_1}{\bar{\Sigma}\Sigma}p_1^\mu-\frac{\bar{\Psi}_2\Psi_2}{\bar{\Sigma}\Sigma}p_2^\mu\right)\sin2\theta\nonumber\\
&-\!\frac{i\hbar m_{I,II}\cos2\theta}{2\!\sqrt{m_1m_2}\bar{\Sigma}\Sigma}\!\left(\nabla^\mu\bar{\Psi}_1\Psi_2\!-\!\bar{\Psi}_1\nabla^\mu\Psi_2\!+\!\nabla^\mu\bar{\Psi}_2\Psi_1\!-\!\bar{\Psi}_2\nabla^\mu\Psi_1\right)\!.
\end{align}
Combing the result (\ref{pIAinp1p2}) with this expression, we immediately check again that at first order in $\hbar$ the 4-momentum $p_{I,II}^\mu$ fulfills the Tulczyjew-M{\o}ller condition in the forms $S_I^{\mu\nu}p_{I,II\nu}=0$ and $S_{II}^{\mu\nu}p_{II,I\nu}=0$. Note also that expression (\ref{pIIIinp1p2}) is, as is expression (\ref{POption3}), symmetric with respect to the substitutions $1\leftrightarrow2$, $\theta\rightarrow-\theta$ and $I\leftrightarrow II$.

From expression (\ref{POption3}), the equation of motion of the 4-momentum $p_{I,II}^{\mu}(x)$ then reads
\begin{widetext}
\begin{equation}\label{DotPOption3}
    \dot{p}_{I,II}^{\mu}=-\frac{1}{2} R^{\mu}_{\;\;\nu\rho\lambda}\varv^\nu\left(\frac{m_{II}\bar{\Phi}_{I}\Phi_{I}}{\sqrt{m_1m_2}\bar{\Sigma}\Sigma}S_{I}^{\rho\lambda}+\frac{m_{I}\bar{\Phi}_{II}\Phi_{II}}{\sqrt{m_1m_2}\bar{\Sigma}\Sigma}S_{II}^{\rho\lambda}\!-\!\hbar m_{I,II}\frac{\bar{\Phi}_{II}\sigma^{\rho\lambda}\Phi_{I}+\bar{\Phi}_{I}\sigma^{\rho\lambda}\Phi_{II}}{2\sqrt{m_1m_2}\bar{\Sigma}\Sigma}\right)_{\mathcal{O}(\hbar)}-(\nabla^\mu \varv_\nu)p_{I,II}^\nu+\mathcal{O}(\hbar^2).
\end{equation}
\end{widetext}
The detailed derivation of this equation is given in Appendix \ref{App8}.

We notice from this last result that, in contrast to the previous expressions (\ref{Dotp}) and (\ref{DotPOption2}) for the dynamical 4-momenta, there emerges here an invariance under the substitutions $1\leftrightarrow2$ and $I\leftrightarrow II$. This symmetry emerges, of course, from the fact that the dynamical 4-momentum (\ref{POption3}) is now also symmetric under the exchange of the two particles of masses $m_I$ and $m_{II}$. Moreover, in the limit $\hbar\rightarrow0$, Eq.\,(\ref{DotPOption3}) reduces to the geodesic equation $\dot{p}_{I,II}^\mu=0$. Note also that, in contrast to what is found for the momenta from the first and the second options, the result (\ref{DotPOption3}) shows that the 4-momentum $p_{I,II}^{\mu}$ does obey an equation that is similar to Eq.\,(\ref{QMPD1}), albeit an extra term similar to the one appearing in the time derivative of the weighted sum $(\bar{\Phi}_I\Phi_Ip_I^{B\mu}+\bar{\Phi}_{II}\Phi_{II}p_{II}^{B\mu})/(\bar{\Phi}_I\Phi_I+\bar{\Phi}_{II}\Phi_{II})$ is also involved here. 

The result (\ref{DotPOption3}) displays explicitly the role of the quantum superposition in dictating the deviation from geodesic motion of the particles. In fact, we observe that both spin tensors $S_I^{\mu\nu}$ and $S_{II}^{\mu\nu}$ associated to the two superposed states $\Phi_I$ and $\Phi_{II}$ contribute now simultaneously to the deviation from geodesic motion via their coupling to the curvature tensor. Such a coupling is weighted by the effective mass of each of the two superpositions $\Phi_I$ and $\Phi_{II}$, respectively. Moreover, there is also an extra coupling with the curvature tensor that involves the `mixed' spin tensor $(\bar{\Phi}_{II}\sigma^{\mu\nu}\Phi_I+\bar{\Phi}_I\sigma^{\mu\nu}\Phi_{II})/(2\bar{\Sigma}\Sigma)$ arising from the interference between the components of the quantum superposition. 

If one proceeds in analogy with what was done for extracting the spin tensors $S_1^{\mu\nu}$ and $S_2^{\mu\nu}$ from the conserved spin current (\ref{GordonDecomp}), then one would obtain from the conserved spin current $J_{sI,II}^\mu(x)$ the `mixed' spin tensor $S_{I,II}^{\mu\nu}=\hbar M\left(\bar{\Sigma}\Gamma^{\mu\nu}{\rm M}^{-1}\Sigma\right)/(2\bar{\Sigma}\Sigma)$. The explicit expression of the latter in terms of the spinors $\Phi_I$ and $\Phi_{II}$ and the spin tensors $S_I^{\mu\nu}$ and $S_{II}^{\mu\nu}$ reads, 
\begin{multline}\label{MixedSpin}
S_{I,II}^{\mu\nu}=\frac{m_{II}\bar{\Phi}_I\Phi_I}{M\bar{\Sigma}\Sigma}S_I^{\mu\nu}+\frac{m_{I}\bar{\Phi}_{II}\Phi_{II}}{M\bar{\Sigma}\Sigma}S_{II}^{\mu\nu}\\
-\frac{\hbar m_{I,II}(\bar{\Phi}_{II}\sigma^{\mu\nu}\Phi_I+\bar{\Phi}_I\sigma^{\mu\nu}\Phi_{II})}{2M\bar{\Sigma}\Sigma}.    
\end{multline}
This tensor satisfies also at first order in $\hbar$ the Tulczyjew-M{\o}ller condition in the forms $S_{I,II}^{\mu\nu}p_{I\nu}=S_{I,II}^{\mu\nu}p_{II\nu}=0$ and $S_{I,II}^{\mu\nu}p_{I,II\nu}=0$. On the other hand, with the help of expressions (\ref{AppEPhi&DotPhi}) and (\ref{App7Phi&DotPhi}), we obtain the equation of motion of this spin tensor up to the first order in $\hbar$ to be: 
\begin{equation}\label{DotSIII}
\dot{S}_{I,II}^{\mu\nu}=0.
\end{equation}
Thus, in contrast to the dynamics of the spin tensors $S_I^{\mu\nu}$ and $S_{II}^{\mu\nu}$ given by Eq.\,(\ref{DotS}), the leading term on the right-hand side of Eq.\,(\ref{DotSIII}) is second-order or higher in $\hbar$. Therefore, just as does the mixed 4-momentum $p_{I,II}^\mu$, this mixed spin tensor obeys the same dynamics as the one obeyed by the spin tensor of the single-state particle.

We conclude from these various observations that no specific option for building the 4-momentum is preferable over the other. Choosing one option over the other depends on one's needs. If one wishes to keep track of one of the two superpositions $\Phi_I$ or $\Phi_{II}$ individually, then the first option for the 4-momentum given by  Eq.\,(\ref{SuperposedDynamicalP}) is the recommended one. If, instead, one is interested in the non-conserved currents $\bar{\Phi}_I\gamma^\mu\Phi_I$ and $\bar{\Phi}_{II}\gamma^\mu\Phi_{II}$ associated to each superposition individually, then one needs to consider the second option for building the dynamical 4-momentum. Finally, if one is rather interested in the conserved total 4-current $\bar{\Phi}_I\gamma^\mu\Phi_I+\bar{\Phi}_{II}\gamma^\mu\Phi_{II}$, one has to rely on the mixed dynamical 4-momentum (\ref{POption3}).   
\section{Application to flavor neutrinos}\label{sec:Application}
In light of our various results concerning superposed states from the previous section, we briefly discuss in this section the case of flavor neutrinos' spin oscillation in curved spacetime. Let, for simplicity, $\ket{\nu_{\rm e}}$ and $\ket{\nu_{\upmu}}$ be the only two flavor states of a neutrino; the electron flavor and the muon flavor, respectively. Let $\Phi_{\rm e}(x)$ and $\Phi_{\upmu}(x)$ be the corresponding spinor fields of the two flavor states, and let $\Psi_1(x)$ and $\Psi_2(x)$ be the corresponding two mass eigenstates' fields of masses $m_1$ and $m_2$, respectively. The flavor states are expandable as linear combinations of the mass eigenstates according to Eq.\,(\ref{PhiI&PhiII}) in which one replaces the subscripts $I$ and $II$ by the flavor subscripts $\rm e$ and $\upmu$, respectively. Furthermore, with such a replacement of subscripts, we also get from Eq.\,(\ref{CoupledDirac}) the coupled Dirac equations satisfied by the flavor states $\Phi_{\rm e}(x)$ and $\Phi_{\upmu}(x)$, and from Eq.\,(\ref{MassesI&II}) we get the masses
$m_{\rm e}$ and $m_{\upmu}$ of the flavor states in terms of the masses $m_1$ and $m_2$ of the mass eigenstates.

Our results from the previous section offer us new tools for complementing the study done in Ref.\,\cite{NSO} on the effect of gravity on the spin oscillation of neutrinos coupled to the scalar field of screening models of dark energy, such as the Chameleon and Symmetron models \cite{Chameleon,JCAP2009,Symmetron,ChameleonTests,ChameleonTests2}. Indeed, the dynamics of the neutrino's spin in the flavor basis $(\Phi_{\rm e},\Phi_{\upmu})$ is given by Eq.\,(\ref{DotS}), up to the first order in $\hbar$ on the right-hand side of the equation, after performing the subscripts substitutions $(I\rightarrow{\rm e},II\rightarrow{\upmu})$. In terms of the spin tensors $S_1^{\mu\nu}$ and $S_2^{\mu\nu}$ of the mass eigenstates $\Psi_1$ and $\Psi_2$
the spin tensor, say of the electron neutrino, reads
\begin{align}
S_{\rm e}^{\mu\nu}&=\frac{\bar{\Psi}_1\Psi_1}{\bar{\Phi}_{\rm e}\Phi_{\rm e}}S_1^{\mu\nu}\cos^2\theta+\frac{\bar{\Psi}_2\Psi_2}{\bar{\Phi}_{\rm e}\Phi_{\rm e}}S_2^{\mu\nu}\sin^2\theta\nonumber\\
&\quad+\frac{\hbar}{4\bar{\Phi}_{\rm e}\Phi_{\rm e}}(\bar{\Psi}_1\sigma^{\mu\nu}\Psi_2+\bar{\Psi}_2\sigma^{\mu\nu}\Psi_1)\sin2\theta,
\end{align}
and its dynamics up to first order in $\hbar$ takes the form:
\begin{multline}\label{DotSe}
\dot{S}_{\rm e}^{\mu\nu}=\left(\frac{i}{\hbar}\frac{\bar{\Psi}_1\Psi_2-\bar{\Psi}_2\Psi_1}{2\bar{\Phi}_{\rm e}\Phi_{\rm e}} S_{\rm e}^{\mu\nu}+i\frac{\bar{\Psi}_2\sigma^{\mu\nu}\Psi_1-\bar{\Psi}_1\sigma^{\mu\nu}\Psi_2}{4\bar{\Phi}_{\rm e}\Phi_{\rm e}}\right)_{\mathcal{O}(\hbar)}\\
\times\Delta m_{21}\sin2\theta+\mathcal{O}(\hbar^2).
\end{multline}
Thus, the dynamics of spin in the flavor basis receives correction terms compared to the equation (\ref{QMPD2}) one obtains for either mass eigenstates separately. These corrections are due to the overlapping of the spinor fields of the mass eigenstates. However, as we see from Eq.\,(\ref{DotSe}), all the corrections in that equation are first-order and higher in $\hbar$. Therefore, if one discards those corrections on the basis that they are higher than the zeroth order in $\hbar$, one simply recovers the approximation made in Ref.\,\cite{NSO} where the second set of the classical MPD equations (\ref{MPD2}) was used after keeping only those terms that are zeroth-order in spin on the right-hand side of that set of equations. Moreover, if one relies instead on the mixed spin tensor $S_{\rm e,\upmu}^{\mu\nu}$ given by Eq.\,(\ref{MixedSpin}), then, at first order in $\hbar$, one falls back on exactly the same spin dynamics one obtains for neutrinos within their mass eigenstates basis. The conclusions reached concerning the effect of the scalar field of screening models within that approximation are therefore not altered.

If, on the other hand, one keeps the corrections on the right-hand side of Eq.\,(\ref{DotSe}) before letting the neutrinos couple to the scalar field, then any extra effect of the scalar field will necessarily be proportional to $\hbar$ as well. To see this, let the scalar field $\varphi(x)$ of such models couple to the neutrino spinor field via the model-dependent regular and nowhere vanishing functional $A(\varphi)$ of the scalar field. The neutrino then propagates within an `effective' spacetime metric $h_{\mu\nu}(x)$ given in terms of the real metric $g_{\mu\nu}(x)$ by $h_{\mu\nu}=A^2(\varphi)g_{\mu\nu}$. In the Chameleon model, the functional $A(\varphi)$ takes the form $A(\varphi) = \exp(\beta \varphi)$ \cite{Chameleon}, whereas in the Symmetron model it takes the form $A(\varphi) = 1 + \beta\varphi^2$, for some arbitrary constant $\beta$ that has the dimensions of an inverse mass. This coupling of the neutrino spinor field with $\varphi(x)$ entails that the effective spacetime vierbeins obtained from the metric $h_{\mu\nu}$ modify the gamma matrices into $\upgamma^\mu=A^{-1}(\varphi)\gamma^\mu$. This, in turn, implies that the coupling of the neutrino to the scalar field simply modifies the right-hand side of Eq.\,(\ref{DotSe}) by the multiplicative factor $A^{-2}(\varphi)$. This shows that, at the leading order, the scalar field's effect on the spin precession is indeed proportional to $\hbar$. Recalling that gravity-induced spin oscillations are already too weak to induce a detectable spin flip on high-energy neutrinos coming from cosmic sources even at the zeroth order in $\hbar$ (as reported in Ref.\,\cite{NSO} and references therein), we conclude that all the corrections from the scalar field $\varphi(x)$ can therefore be safely neglected given that they are all proportional to $\hbar$.

To discuss the equation of motion of the dynamical 4-momentum of these neutrinos and their deviation from geodesic motion, we need, in light of our discussion in the last paragraph below Eq.\,(\ref{DotPOption3}), to rely on one of the three options depending on which feature of the superposition one is interested in. For this reason, we shall in what follows consider all three different options for the coupled neutrinos. 

Using our results (\ref{Dotp}), (\ref{DotPOption2}) and (\ref{DotPOption3}) and the definition $m_{\rm e\upmu}=\frac{1}{2}\Delta m_{21}\sin^2\theta$ of the coupling mass between the flavor neutrinos, we learn that, up to first order in $\hbar$, the corrections brought by the quantum superposition of the mass eigenstates to the first set of equations governing the dynamics of both neutrino flavors $\ket{\nu_{\rm e}}$ and $\ket{\nu_{\upmu}}$ are all due to the interference between the mass eigenstates. Only the weighted sums $(\bar{\Phi}_{\rm e}\Phi_{\rm e}p_{\rm e}^{A\mu}+\bar{\Phi}_{\upmu}\Phi_{\upmu}p_{\upmu}^{A\mu})/(\bar{\Phi}_{\rm e}\Phi_{\rm e}+\bar{\Phi}_{\upmu}\Phi_{\upmu})$ and $(\bar{\Phi}_{\rm e}\Phi_{\rm e}p_{\rm e}^{B\mu}+\bar{\Phi}_{\upmu}\Phi_{\upmu}p_{\upmu}^{B\mu})/(\bar{\Phi}_{\rm e}\Phi_{\rm e}+\bar{\Phi}_{\upmu}\Phi_{\upmu})$ as well as the mixed 4-momentum $p_{\rm e,\upmu}^{\mu}$ obey an equation similar to the one obeyed by the mass eigenstates individually.
Moreover, in contrast to neutrino flavor oscillation, which involves only the difference of the masses squared of the mass eigenstates, Eqs.\,(\ref{Dotp}), (\ref{DotPOption2}) and (\ref{DotPOption3}) imply that deviation from geodesic motion due to the superposed nature of neutrinos involves mass differences, the product of the masses $m_1m_2$ as well as the product $m_{\rm e}m_{\upmu}$ of the flavor states' masses and the squared coupling mass $m_{\rm e,\upmu}^2$.

Now, this coupling of the neutrinos with the scalar field $\varphi(x)$ entails that the neutrinos proper time element $\rm d \tau$ is modified into ${\rm d}\uptau = A(\varphi){\rm d}\tau$, yielding an effective 4-velocity $\varv^\mu$ along the geodesic path given by $\varv^\mu=A^{-1}(\varphi)\varv^\mu$ \cite{NSO}. Similarly, the effective spacetime vierbeins obtained from the metric $h_{\mu\nu}$ modify the gamma matrices into $\upgamma^\mu=A^{-1}(\varphi)\gamma^\mu$ and the spin connection is modified into $\upomega_\mu^{ab}=\omega_\mu^{\,\,ab}-A_{,\nu}\left(e^{\nu a}e^b_\mu-e^{\nu b}e^a_\mu\right)/A$ \cite{NSO}, where $A_{,\mu}$ denotes the partial derivative of $\partial_\mu A(\varphi)$ with respect to the coordinate $x^\mu$. 
The spin tensor of the coupled neutrinos is given in terms of the spin tensors $S_{\rm e}^{\mu\nu}$ and $S_{\upmu}^{\mu\nu}$ of the uncoupled neutrinos by $A^{-2}(\varphi)S_{\rm e}^{\mu\nu}$ and $A^{-2}(\varphi)S_{\upmu}^{\mu\nu}$, respectively \cite{NSO}. Moreover, one easily shows that replacing the spacetime metric by the effective metric $h_{\mu\nu}=A^2(\varphi)g_{\mu\nu}$ seen by the coupled neutrinos, the curvature tensor $\mathcal{R}_{\mu\nu\rho\lambda}$ seen by the coupled neutrinos gets also modified to \cite{Wald}:
\begin{align}
\mathcal{R}^\mu_{\;\;\nu\rho\lambda}&=R^\mu_{\;\;\nu\rho\lambda}+\delta^\mu_{\lambda}\left(\frac{A_{;\nu\rho}}{A}-2\frac{A_{,\nu}A_{,\rho}}{A^2}\right)-g_{\nu\lambda}\left(\frac{A^{,\mu}_{\;\;;\rho}}{A}-2\frac{A_{,\rho}A^{,\mu}}{A^2}\right)\nonumber\\
&\quad-\delta^\mu_\rho\left(\frac{A_{;\nu\lambda}}{A}-2\frac{A_{,\nu}A_{,\lambda}}{A^2}+g_{\nu\lambda}\frac{A_{,\alpha}A^{,\alpha}}{A^2}\right)\nonumber\\
&\quad+g_{\nu\rho}\left(\frac{A^{,\mu}_{\;\;;\lambda}}{A}-2\frac{A^{,\mu}A_{,\lambda}}{A^2}+\delta^\mu_\lambda\frac{A_{,\alpha}A^{,\alpha}}{A^2}\right).
\end{align}
A semi-colon denotes a covariant derivative. The corresponding Christoffel symbols $\Gamma^{\mu}_{\nu\rho}$ become $\Upgamma^{\mu}_{\nu\rho}=\Gamma^\mu_{\nu\rho}+(\delta^\mu_\nu A_{,\rho}+\delta^\mu_\rho A_{,\nu}-g_{\nu\rho}A^{,\mu})/A$. Plugging these various identities into Eqs.\,(\ref{Dotp}), (\ref{DotPOption2}) and (\ref{DotPOption3}), and then taking account of the fact that $\varv_\mu S_{\rm e}^{\mu\nu}=\varv_\mu S_{\upmu}^{\mu\nu}=0$, leads, at first order in $\hbar$, to the following equations of motion for the three possible dynamical 4-momenta of coupled neutrinos:
\begin{widetext}
\begin{align}\label{DotPCoupledNeutrinos}
\dot{p}_{\rm e}^{A\mu}&=-\frac{1}{2A(\varphi)} R^{\mu}_{\;\;\nu\rho\lambda}\varv^\nu S_{\rm e}^{\rho\lambda}-(\nabla^\mu \varv_{\nu})p_{\rm e}^{A\nu}+\left(\frac{iA(\varphi)}{\hbar}\frac{\bar{\Psi}_1\Psi_2-\bar{\Psi}_2\Psi_1}{2\bar{\Phi}_{\rm e}\Phi_{\rm e}}\, p_{\rm e}^{A\mu}+\frac{\nabla^\mu\bar{\Psi}_1\Psi_2-\bar{\Psi}_1\nabla^\mu\Psi_2+\bar{\Psi}_2\nabla^\mu\Psi_1-\nabla^\mu\bar{\Psi}_2\Psi_1}{4\bar{\Phi}_{\rm e}\Phi_{\rm e}}\right)_{\mathcal{O}(\hbar)}\Delta m_{21}\sin2\theta\nonumber\\
&\quad+\mathcal{O}(\hbar^2),\\
    \dot{p}_{\rm e}^{B\mu}&=-\frac{m_{\rm e} m_{\upmu}}{2(m_{\rm e}m_{\upmu}-m^2_{{\rm e},\upmu})A(\varphi)}R^{\mu}_{\;\;\nu\rho\lambda}\varv^\nu\left(S_{\rm e}^{\rho\lambda}-\frac{\hbar m_{{\rm e},\upmu}}{m_{\upmu}}\frac{\bar{\Phi}_{\upmu}\sigma^{\rho\lambda}\Phi_{\rm e}+\bar{\Phi}_{\rm e}\sigma^{\rho\lambda}\Phi_{\upmu}}{4\bar{\Phi}_{\rm e}\Phi_{\rm e}}\right)_{\mathcal{O}(\hbar)}-(\nabla^\mu \varv_{\nu})p_{\rm e}^{B\nu}\nonumber\\
    &\quad+\left(\frac{m_{\rm e} m_{\upmu}}{m_{\rm e}m_{\upmu}-m^2_{{\rm e},\upmu}}\frac{\nabla^\mu\bar{\Psi}_1\Psi_2-\bar{\Psi}_1\nabla^\mu\Psi_2+\bar{\Psi}_2\nabla^\mu\Psi_1-\nabla^\mu\bar{\Psi}_2\Psi_1}{4\bar{\Phi}_{\rm e}\Phi_{\rm e}}+\frac{iA(\varphi)}{\hbar}\frac{\bar{\Psi}_1\Psi_2-\bar{\Psi}_2\Psi_1}{2\bar{\Phi}_{\rm e}\Phi_{\rm e}}p_{\rm e}^{B\mu}\right)_{\mathcal{O}(\hbar)}\Delta m_{21}\sin2\theta\nonumber\\
    &\quad+\frac{m_{\rm e} m_{{\rm e},\upmu}\Delta m_{21}}{m_{\rm e}m_{\upmu}-m^2_{{\rm e},\upmu}}\left(\frac{\nabla^\mu\bar{\Psi}_2\Psi_1+\bar{\Psi}_1\nabla^\mu\Psi_2}{2\bar{\Phi}_{\rm e}\Phi_{\rm e}}\cos^2\theta+\frac{\nabla^\mu\bar{\Psi}_1\Psi_2+\bar{\Psi}_2\nabla^\mu\Psi_1}{2\bar{\Phi}_{\rm e}\Phi_{\rm e}}\sin^2\theta\right)_{\mathcal{O}(\hbar)}+\mathcal{O}(\hbar^2),\\
    \dot{p}_{{\rm e},\upmu}^{\mu}&=-\frac{1}{2A(\varphi)} R^{\mu}_{\;\;\nu\rho\lambda}\varv^\nu\left(\frac{m_{\upmu}\bar{\Phi}_{\rm e}\Phi_{\rm e}}{\sqrt{m_1m_2}\bar{\Sigma}\Sigma}S_{\rm e}^{\rho\lambda}+\frac{m_{\rm e}\bar{\Phi}_{\upmu}\Phi_{\upmu}}{\sqrt{m_1m_2}\bar{\Sigma}\Sigma}S_{\upmu}^{\rho\lambda}-\hbar m_{\rm e,\upmu}\frac{\bar{\Phi}_{\upmu}\sigma^{\rho\lambda}\Phi_{\rm e}+\bar{\Phi}_{\rm e}\sigma^{\rho\lambda}\Phi_{\upmu}}{2\sqrt{m_1m_2}\bar{\Sigma}\Sigma}\right)_{\mathcal{O}(\hbar)}-(\nabla^\mu \varv_\nu) p_{{\rm e},\upmu}^\nu\nonumber\\
    &\quad+\frac{A_{,\nu}(\varphi)}{A(\varphi)}\left(\varv^\mu p_{{\rm e},\upmu}^\nu-p_{{\rm e},\upmu}^\mu \varv^\nu \right)+\mathcal{O}(\hbar^2).
\end{align}
\end{widetext}
These results show that the leading terms in the corrections brought to the equations of motion of coupled neutrinos are first-order in $\hbar$ only for the dynamical 4-momentum $p_{\rm e,\upmu}^\mu$. Indeed, all the terms, including the additional term proportional to $\varv^\mu p^\nu-p^\mu \varv^\nu$, in the equation giving $\dot{p}_{\rm e,\upmu}^\mu$ vanish at the zeroth order in $\hbar$, whereas the leading interference term that is proportional to the coupling factor $A(\varphi)$ in $\dot{p}_{\rm e}^\mu$ and $\dot{p}_{\upmu}^\mu$ is zeroth-order in $\hbar$. Therefore, approximating at the zeroth order in $\hbar$ the equation of motion of the 4-momentum of coupled neutrinos when taking into account the superposed nature of the latter by the geodesic equation obeyed by the mass eigenstates remains valid only within the third-option 4-momentum $p_{\rm e,\upmu}^\mu$. In other words, only based on such a prescription for a dynamical 4-momentum does working either within the flavor basis or within the mass eigenstates basis not affect the outcome for neutrinos coupled to the scalar field of screening models such as the Chameleon and the Symmetron at the zeroth order in $\hbar$.
\section{Summary and discussion}\label{sec:Conclusion}
We have examined in this paper the dynamics of spin-$\frac{1}{2}$ particles freely propagating in curved spacetime by extracting MPD-like equations from a WKB approximation of the Dirac equation. After building the necessary tools and identities for tackling the case of superposed states within the WKB approach, we made use of our results to extract the dynamical equations for particles made of a superposition of two different states, each carrying a different mass. We found that both the equation of motion of the 4-momentum and the dynamics of the spin tensor of such particles receive corrections that are all coming from the interference of the individual spinor fields of the superpositions. This interference comes not only from a direct overlap of the fields, but also from a gravity-induced overlap thanks to the curved spacetime. 

In order to investigate the equation of motion of the dynamical 4-momentum that should go into MPD-like equations for particles propagating as a superposition of states, we explored three different ways of building such a 4-momentum. We denoted those 4-momenta from each option by ($p_{I}^{A\mu}$, $p_{II}^{A\mu}$), ($p_{I}^{B\mu}$, $p_{II}^{B\mu}$) and $p_{I,II}^{\mu}$. We showed that the formulation among those three options that leads to the geodesic equation in the limit $\hbar\rightarrow0$ is the 4-momentum $p_{I,II}^{\mu}$ extracted from the Gordon decomposition of the conserved total 4-current $\bar{\Phi}_I\gamma^\mu\Phi_I+\bar{\Phi}_{II}\gamma^\mu\Phi_{II}$. We found that this is not the case, neither for the pair ($p_{I}^{A\mu}$, $p_{II}^{A\mu}$), obtained by replacing $\Psi_1$ and $\Psi_2$ by $\Phi_I$ and $\Phi_{II}$ in the expressions of $p_1^\mu$ and $p_2^\mu$ of single-state particles, nor for the pair ($p_{I}^{B\mu}$, $p_{II}^{B\mu}$), extracted from the Gordon decomposition of the non-conserved currents $\bar{\Phi}_I\gamma^\mu\Phi_I$ and $\bar{\Phi}_{II}\gamma^\mu\Phi_{II}$ associated separately to each particle. Moreover, we found that only the 4-momentum $p_{I,II}^{\mu}$ obeys an equation similar to the first set of classical  MPD equations. The dynamics of the two pairs $(p_I^{A\mu},p_{II}^{A\mu})$ and $(p_I^{B\mu},p_{II}^{B\mu})$ displays departures from the MPD equations that are due to interference terms between the spinors of the superposition. We, nevertheless, found that the weighted sums $(\bar{\Phi}_I\Phi_Ip_I^{A\mu}+\bar{\Phi}_{II}\Phi_{II}p_{II}^{A\mu})/(\bar{\Phi}_I\Phi_I+\bar{\Phi}_{II}\Phi_{II})$ and $(\bar{\Phi}_{I}\Phi_{I}p_I^{B\mu}+\bar{\Phi}_{II}\Phi_{II}p_{II}^{B\mu})/(\bar{\Phi}_I\Phi_I+\bar{\Phi}_{II}\Phi_{II})$ do obey an MPD-like dynamics equation, with the weighted spin tensor $(\bar{\Phi}_I\Phi_IS_I^{\mu\nu}+\bar{\Phi}_{II}\Phi_{II}S_{II}^{\mu\nu})/(\bar{\Phi}_I\Phi_I+\bar{\Phi}_{II}\Phi_{II})$ playing the role of an effective spin tensor.

Concerning the spin tensor, we showed that no such distinction is necessary as the definition of such a tensor leads to separately conserved spin currents for each particle. Within all three options, we found that in the absence of a mass difference between the different states making the superposition the results reduce to those of the well-known single-state particles. In addition, we found that the proper time derivative of the weighted sum $(\bar{\Phi}_I\Phi_IS_I^{\mu\nu}+\bar{\Phi}_{II}\Phi_{II}S_{II}^{\mu\nu})/(\bar{\Phi}_I\Phi_I+\bar{\Phi}_{II}\Phi_{II})$ of the spin tensors vanishes at first order in $\hbar$, and that, as in the case of single-state particles, the spin tensors $S_I^{\mu\nu}$ and $S_{II}^{\mu\nu}$ satisfy individually the Tulczyjew-M{\o}ller supplementary condition with all three different 4-momenta: $S_r^{\mu\nu}p^{A}_{s\nu}=S_r^{\mu\nu}p^{B}_{s\nu}=S_r^{\mu\nu}p_{I,II\nu}=0$, for $r,s=I,II$. A mixed spin tensor $S_{I,II}^{\mu\nu}$ has also been extracted from the conserved total spin current. Its dynamics is found to be exactly the same at the leading order in $\hbar$ as that of the spin tensor $S_i^{\mu\nu}$ of single-state particles.

We also pointed out that the MPD-like equations obtained from the WKB approximation of the Dirac equation, both for single-state and for multi-state particles, are slightly different from the classical MPD equations. All the equations of motion of the dynamical 4-momentum contain extra terms on their right-hand side that are not present in the first set of classical MPD equations. The equation describing the spin dynamics, on the other hand, lacks terms that are present in the second set of classical MPD equations. We remarked that such a distinction does not arise within the Lagrangian approach \cite{Pavsic,Barut,GaioliGarcia,NuriUnal} as opposed to a WKB approach. One proposal towards understanding this distinction can be found in Ref.\,\cite{GaioliGarcia}. We believe that this point requires more investigation, though, which we shall leave for a future work.

We then argued that the immediate and most natural application of our present results can be made in the field of neutrino physics. Indeed, neutrinos are not only spin-$\frac{1}{2}$ particles, but are also flavor particles made of a superposition of different states having different masses; the so-called mass eigenstates. In light of our general results, we argued that the interference of the mass eigenstates do not bring large modifications to neutrino dynamics in curved spacetime since the leading terms of such modifications all consist of first-order terms in $\hbar$. In fact, given that both spin dynamics and the equation of motion of neutrinos' 4-momentum in the flavor basis receive correction terms that are all first-order and higher in $\hbar$, restricting the latter dynamics to the zeroth order in spin as it is done so far in the literature is amply sufficient for many low-curvature applications. We, nevertheless,  pointed out that within the first two prescriptions $p_{\rm e}^{A\mu}$ and $p_{\rm e}^{B\mu}$ for the dynamical 4-momentum, zeroth-order terms in $\hbar$ do arise. Similarly, the dynamics of neutrinos coupled to the scalar field of screening models is shown to be not affected at the zeroth order in $\hbar$ by the superposed nature of neutrinos only within the third prescription $p_{\rm e,\upmu}^\mu$ for the dynamical 4-momentum.

Although our study was done here for the case of two-flavor neutrinos, a generalization to the case of three-flavor neutrinos should not bring in any particular conceptual difficulties. Also, since our study here has been general enough, adapting our present results to the case of neutrinos propagating inside matter using more quantum field theoretical tools (see e.g., Ref.\,\cite{LambiasePapini,NSOMatter} and references therein) should not present in principle extra challenges. However, a proper investigation of such a problem  will be left for future works as well.
\section*{Acknowledgments}
The authors are grateful to the anonymous referee for the constructive comments that greatly helped improve our manuscript. F.\,H., M.\,S. and R.\,S. are supported by the Natural Sciences and Engineering Research Council of Canada (NSERC) Discovery Grant No. RGPIN-2017-05388, and by the Fonds de Recherche du Québec - Nature et Technologies (FRQNT). A.\,L. is supported by an Atlantic Association for Research in the Mathematical Sciences (AARMS) Fellowship.

\appendix
\section{Derivation of the constraints (\ref{DotThetaA&DotThetaB})}\label{App1}
The constraints (\ref{DotThetaA&DotThetaB}) can be derived as follows. Using the fact that $\Theta_{A}(x)$ and $\Theta_{B}$ are solutions to Eq.\,(\ref{0thOrder}), we have $(\gamma^\mu \varv_{\mu}+\mathbb{I})\Theta_{A}=0$ and $(\gamma^\mu \varv_{\mu}+\mathbb{I})\Theta_{B}=0$. Next, applying the proper time derivative operator ${\rm d}/{\rm d}\tau$ to both sides of these two identities and taking account of the geodesic equation ${\rm D}\varv^\mu/{{\rm d}\tau}=0$, yields
\begin{equation}
\left(\gamma^\mu \varv_{\mu}+\mathbb{I}\right)\dot{\Theta}_{A}=0,\qquad
     \left(\gamma^\mu \varv_{\mu}+\mathbb{I}\right)\dot{\Theta}_B=0.
\end{equation}
These two identities show that $\dot{\Theta}_{A}$ and $\dot{\Theta}_{B}$ are, each, again a solution to the homogeneous equation (\ref{0thOrder}). Therefore, $\dot{\Theta}_{A}$ and $\dot{\Theta}_{B}$ can be written as the following two linear combinations:
\begin{equation}\label{App1DotThetaA&DotThetaB}
\dot{\Theta}_A=C_{1}\Theta_{A}+C_{2}\Theta_{B},\qquad
\dot{\Theta}_{B}=D_{1}\Theta_{B}+D_{2}\Theta_{A},
\end{equation}
for some four arbitrary complex scalars $C_{1}$, $C_{2}$, $D_{1}$, $D_{2}$. These four scalars must, in turn, satisfy certain conditions. Indeed, applying the derivative operator ${\rm d}/{\rm d}\tau$ to the normalization conditions $\bar{\Theta}_{A}\Theta_{A}=1$ and $\bar{\Theta}_{B}\Theta_{B}=1$, respectively, implies, after using Eq.\,(\ref{App1DotThetaA&DotThetaB}) as well as the orthogonality condition $\bar{\Theta}_{A}\Theta_{B}=0$, that $C_{1}^*=-C_{1}$ and $D_{1}^*=-D_{1}$. On the other hand, applying the operator ${\rm d}/{\rm d}\tau$ to the orthogonality condition $\bar{\Theta}_{A}\Theta_{B}=0$, implies that $C_2^*=-D_2$. 
\section{Derivation of Eq.\,(\ref{BisThetaA&B})}\label{App2}
The constraints (\ref{BisThetaA&B}) are derived as follows:
\begin{align}\label{App2BisThetaA&B}
\bar{\Theta}_{A}\gamma^\mu\nabla_\mu\Theta_{A}&=-\bar{\Theta}_A\gamma^\nu\nabla_\nu(\gamma^\mu \varv_\mu\Theta_A)\nonumber\\
&=-\bar{\Theta}_A\gamma^\nu\gamma^\mu(\nabla_\nu \varv_\mu)\Theta_A-\bar{\Theta}_A\gamma^\nu\gamma^\mu \varv_\mu\nabla_\nu\Theta_A\nonumber\\
&=\nabla_\mu \varv^\mu+2\bar{\Theta}_A \varv^\mu\nabla_\mu\Theta_A+\bar{\Theta}_A\gamma^\mu\gamma^\nu \varv_\mu\nabla_\nu\Theta_A\nonumber\\
&=\nabla_\mu \varv^\mu+2C_1-\bar{\Theta}_A\gamma^\mu\nabla_\mu\Theta_A\nonumber\\
&=\tfrac{1}{2}\nabla_\mu \varv^\mu+C_1.
\end{align}
In the third step, we used $\nabla_\nu \varv_\mu=\nabla_\mu \varv_\nu$ and $\gamma^\mu\gamma^\nu=-2g^{\mu\nu}-\gamma^\nu\gamma^\mu$. In the fourth step, we used the first identity in Eq.\,(\ref{App1DotThetaA&DotThetaB}). Next, replacing $\Theta_A$ by $\Theta_B$ everywhere in the derivation (\ref{App2BisThetaA&B}) simply turns $C_1$ into $D_1$ in the fourth step, proving thus the second identity in Eq.\,(\ref{BisThetaA&B}). On the other hand, keeping $\bar{\Theta}_A$ in this derivation and replacing only the $\Theta_A$ on the right by $\Theta_B$ removes the term $\nabla_\mu \varv^\mu$ from the third line and turns $C_1$ into $D_2$ in the fourth line thanks to the second identity in Eq.\,(\ref{App1DotThetaA&DotThetaB}), so that we end up with $\bar{\Theta}_A\gamma^\mu\nabla_\mu\Theta_B=D_2$. 
\section{Derivation of Eq.\,(\ref{psi0Dotpsi1Andpsi1Dotpsi0})}\label{App3}
We derive here Eq.\,(\ref{psi0Dotpsi1Andpsi1Dotpsi0}) by setting $n=1$ in Eq.\,(\ref{Dotpsi(n)}) and multiplying both sides of the equation from the left by $\bar{\psi}_i^{(0)}$ to get,
\begin{align}\label{C1}
\bar{\psi}_i^{(0)}\dot{\psi}_j^{(1)}+\tfrac{1}{2}(\nabla_\mu \varv^\mu)\bar{\psi}_i^{(0)}\psi_j^{(1)}&=-\bar{\psi}_i^{(0)}\gamma^\mu\nabla_\mu\xi_j^{(1)}\nonumber\\
&=-\frac{i}{2m_j}\bar{\psi}_i^{(0)}\gamma^\mu\gamma^\nu\nabla_\mu\nabla_\nu\psi_j^{(0)}.
\end{align}
In the second line we used Eq.\,(\ref{XiSpinor}) with $n=1$, the completeness relation $\sum_{s=A,B}(\Theta_{is}\bar{\Theta}_{is}-\Pi_{is}\bar{\Pi}_{is})=1$ and the conditions $\bar{\Theta}_{A}\gamma^\mu\nabla_\mu\psi_i^{(0)}=\bar{\Theta}_{B}\gamma^\mu\nabla_\mu\psi_i^{(0)}=0$ that allowed us to turn the term $-\bar{\psi}_i^{(0)}\gamma^\mu\nabla_\mu\xi^{(1)}_j$ into $-\frac{i}{2m_j}\bar{\psi}_i^{(0)}\gamma^\mu\gamma^\nu\nabla_\mu\nabla_\nu\psi_j^{(0)}$.
As the left-hand side of Eq.\,(\ref{C1}) vanishes in Riemann normal coordinates in the comoving reference frame of the particle, it follows that the term on the right-hand side, being a scalar and a relativistic invariant that is independent of the four-velocity $\varv^\mu$, should also vanish identically in all reference frames. Therefore, we conclude that in all reference frames we have
\begin{align}
\bar{\psi}_i^{(0)}\dot{\psi}_j^{(1)}&=-\tfrac{1}{2}(\nabla_\mu \varv^\mu)\bar{\psi}_i^{(0)}\psi_j^{(1)},\nonumber\\
\dot{\bar{\psi}}_i^{(1)}\psi_j^{(0)}&=-\tfrac{1}{2}(\nabla_\mu \varv^\mu)\bar{\psi}_i^{(1)}\psi_j^{(0)}.
\end{align}
The second identity is obtained by taking the complex conjugate of the first identity and  switching around the indices $i$ and $j$.


\section{Derivation of Eq.\,(\ref{QMPD1})}\label{App4}
We provide in this appendix a detailed derivation of Eq.\,(\ref{QMPD1}). As stated in the Introduction, the derivation we give here is very well suited for applying it to the case of superposed states as it consists of a simple series of rearrangements combined with the various identities derived in a very general way in Sec.\,\ref{sec:MPDfromWKB}. 

First, according to the definition (\ref{QuantumDynamical4-momentum}) of the dynamical 4-momentum $p_i^\mu$ in terms of the zeroth-order 4-spinor $\psi_i^{(0)}(x)$, we have
\begin{align}\label{AppBFirstStep}
\dot{p}_i^\mu&=\dot{\pi}_i^\mu+i\hbar\frac{\rm d}{{\rm d}\tau} \left[\frac{\nabla^\mu\bar{\psi}_i^{(0)}\psi_i^{(0)}-\bar{\psi}_i^{(0)}\nabla^\mu\psi_i^{(0)}}{2\bar{\psi}_i^{(0)}\psi_i^{(0)}}\right]\nonumber\\
    &=(\nabla_\nu \varv^\nu)(p_i^\mu-\pi_i^\mu)+i\hbar \varv_{\nu} \frac{\nabla^\nu\nabla^\mu\bar{\psi}_i^{(0)}\psi_i^{(0)}-\bar{\psi}_i^{(0)}\nabla^\nu\nabla^\mu\psi_i^{(0)}}{2\bar{\psi}_i^{(0)}\psi_i^{(0)}}\nonumber\\
    &\quad+i\hbar \frac{\nabla^\mu\bar{\psi}_i^{(0)}\dot{\psi}_i^{(0)}-\dot{\bar{\psi}}_i^{(0)}\nabla^\mu\psi_i^{(0)}}{2\bar{\psi}_i^{(0)}\psi_i^{(0)}}.
\end{align}
In the second line, we used the geodesic equation $\dot{\pi}_i^\mu=0$, and then we made use of Eq.\,(\ref{Dotpsi(o)}), as well as the fact that ${\rm d}/{\rm d}\tau=\varv_\mu\nabla^\mu$.

Next, by switching the order of the covariant derivatives in the numerator of the second term in the second line on the right-hand side of Eq.\,(\ref{AppBFirstStep}) and using the fact that (see e.g., Ref.\,\cite{SchroDirac}) $[\nabla_\mu,\nabla_\nu]\psi_i^{(0)}=\frac{i}{4}R_{\mu\nu ab}\sigma^{ab}\psi_i^{(0)}$ and $[\nabla_\mu,\nabla_\nu]\bar{\psi}_i^{(0)}=-\frac{i}{4}R_{\mu\nu ab}\bar{\psi}_i^{(0)}\sigma^{ab}$, together with the definition (\ref{SpinTensor}) of the spin tensor $S_i^{\mu\nu}$ to first order in $\hbar$, Eq.\,(\ref{AppBFirstStep}) takes the following form:
\begin{align}\label{AppBSecondStep}
    \dot{p}_i^\mu&=(\nabla_\nu \varv^\nu)(p_i^\mu-\pi_i^\mu)-\frac{1}{2}R^{\mu}_{\;\;\nu\rho\sigma} \varv^\nu S_i^{\rho\sigma}\nonumber\\
    &\quad+i\hbar \nabla^\mu\left[\frac{\dot{\bar{\psi}}_i^{(0)}\psi_i^{(0)}-\bar{\psi}_i^{(0)} \dot{\psi}_i^{(0)}}{2\bar{\psi}_i^{(0)}\psi_i^{(0)}}\right]\nonumber\\
&\quad+i\hbar\frac{\nabla^\mu[\bar{\psi}_i^{(0)}\psi_i^{(0)}]}{\bar{\psi}_i^{(0)}\psi_i^{(0)}}\frac{\dot{\bar{\psi}}_i^{(0)}\psi_i^{(0)}-\bar{\psi}_i^{(0)}\dot{\psi}_i^{(0)}}{2\bar{\psi}_i^{(0)}\psi_i^{(0)}}\nonumber\\
&\quad-(\nabla^\mu \varv_\nu)p_i^\nu+i\hbar\frac{\nabla^\mu\bar{\psi}_i^{(0)} \dot{\psi}_i^{(0)}-\dot{\bar{\psi}}_i^{(0)}\nabla^\mu\psi_i^{(0)}}{\bar{\psi}_i^{(0)}\psi_i^{(0)}}.
    \end{align}
Upon using Eqs.\,(\ref{Dotpsi(o)}) and (\ref{QuantumDynamical4-momentum}), Eq.\,(\ref{AppBSecondStep}) takes the following final form:
\begin{equation}\label{AppBQMPD1}
\dot{p}_i^\mu=-\tfrac{1}{2}R^\mu_{\;\;\nu\rho\sigma}\varv^\nu S_i^{\rho\sigma}-(\nabla^\mu \varv_{\nu}) p_i^\nu.  
\end{equation}

\section{Derivation of Eq.\,(\ref{Dotp})}\label{App5}
We give in this appendix the detailed steps leading to Eq.\,(\ref{Dotp}). Starting from the definition (\ref{SuperposedDynamicalP}) of the dynamical 4-momentum $p_I^\mu(x)$, carried by the spinor field $\Phi_I(x)$, we first compute the proper time derivative $\dot{p}_I^\mu(x)$ by performing the same rearrangements of terms we performed to extract Eqs.\,(\ref{AppBFirstStep}) and (\ref{AppBSecondStep}). The result is:
\begin{align}\label{AppEFirstStepDotP}
    \dot{p}_I^{A\mu}&=i\hbar \frac{\rm d}{{\rm d}\tau} \left(\frac{\nabla^\mu\bar{\Phi}_I\Phi_I-\bar{\Phi}_I\nabla^\mu\Phi_I}{2\bar{\Phi}_I\Phi_I}\right)\nonumber\\
    &=-\frac{\overset{\boldsymbol .}{(\bar{\Phi}_I\Phi_I)}}{\bar{\Phi}_I\Phi_I}p_I^{A\mu}-\frac{1}{2} R^{\mu}_{\;\;\nu\rho\lambda}\varv^\nu S_I^{\rho\lambda}+i\hbar \frac{\nabla^\mu(\dot{\bar{\Phi}}_I\Phi_I-\bar{\Phi}_I\dot{\Phi}_I)}{2\bar{\Phi}_I\Phi_I}\nonumber\\
    &\quad-(\nabla^\mu \varv_{\nu})p_I^{A\nu}+i\hbar \frac{\nabla^\mu\bar{\Phi}_I\dot{\Phi}_I-\dot{\bar{\Phi}}_I\nabla^\mu\Phi_I}{\bar{\Phi}_I\Phi_I}.
\end{align}
The next step is to compute, up to the leading order in $\hbar$, the derivatives $\dot{\Phi}_I$ and $\nabla^\mu\Phi_I$ and the various products involving the latter in Eq.\,(\ref{AppEFirstStepDotP}).

We first use expression (\ref{PhiI&PhiII}) of the spinor field $\Phi_I(x)$ in terms of the component spinor fields $\Psi_1(x)$ and $\Psi_2(x)$ for which we already found in Sec.\,\ref{sec:MPDfromWKB} the various relevant identities satisfied by the spinors $\psi_i^{(0)}(x)$ and $\psi_i^{(1)}(x)$ in their WKB expansions. Keeping only terms up to the leading orders in $\hbar$ in each, we have the following expansions of $\Phi_I(x)$ and $\dot{\Phi}_I(x)$, respectively:
\begin{align}\label{AppEPhi&DotPhi}
\Phi_I&=e^{\frac{i}{\hbar}\mathcal{S}_1}\left(\psi_1^{(0)}+\hbar\psi_1^{(1)}+\hbar^2\psi_1^{(2)}+\dots\right)\cos\theta\nonumber\\
&\quad+e^{\frac{i}{\hbar}\mathcal{S}_2}\left(\psi_2^{(0)}+\hbar\psi_2^{(1)}+\hbar^2\psi_2^{(2)}+\ldots\right)\sin\theta,\nonumber\\
\dot{\Phi}_I&=e^{\frac{i}{\hbar}\mathcal{S}_1}\left[\dot{\psi}_1^{(0)}+\hbar\dot{\psi}_1^{(1)}+\hbar^2\dot{\psi}_1^{(2)}+\ldots\right.\nonumber\\
&\quad\left.-\tfrac{i}{\hbar}m_1(\psi_1^{(0)}+\hbar\psi_1^{(1)}+\hbar^2\psi_1^{(2)}+\ldots)\right]\cos\theta\nonumber\\
&\quad+e^{\frac{i}{\hbar}\mathcal{S}_2}\left[\dot{\psi}_2^{(0)}+\hbar\dot{\psi}_2^{(1)}+\hbar^2\dot{\psi}_2^{(2)}+\ldots\right.\nonumber\\
&\quad\left.-\tfrac{i}{\hbar}m_2\left(\psi_2^{(0)}+\hbar\psi_2^{(1)}+\hbar^2\psi_2^{(2)}+\ldots\right)\right]\sin\theta.
\end{align}
The expression of $\dot{\Phi}_I(x)$ is obtained by making use of the identity $\dot{\mathcal{S}}_i=-m_i$. Using these results, and making use of Eqs.\,(\ref{Dotpsi(o)}) and (\ref{psi0Dotpsi1Andpsi1Dotpsi0}), we compute the proper time derivative $\overset{\boldsymbol .}{(\bar{\Phi}_I\Phi_I)}$ to be:
\begin{align}\label{DotPhiPhi}
\overset{\boldsymbol .}{(\bar{\Phi}_I\Phi_I)}&=\left[-\bar{\Phi}_I\Phi_I\,(\nabla_\mu \varv^\mu)\right.\nonumber\\
&\;\quad\left.+\frac{i}{\hbar}\left(\bar{\Psi}_2\Psi_1-\bar{\Psi}_1\Psi_2\right)\Delta m_{21}\cos\theta\sin\theta\right]_{\mathcal{O}(\hbar)}+\mathcal{O}(\hbar^2).
\end{align}
As indicated in Sec.\,\ref{sec:SuperposedMPD}, a subscript $\mathcal{O}(\hbar)$ on brackets means that we take only those terms inside the brackets that are at most of order $\hbar$. Also, $\Delta m_{21}$ stands for the difference $m_2-m_1$. In a similar manner, we compute the numerators in the third term and in the last term on the right-hand side of Eq.\,(\ref{AppEFirstStepDotP}), up to first order in $\hbar$, to be, respectively:
\begin{align}\label{AppEIntermediate1}
&i\hbar\nabla^\mu\left(\dot{\bar{\Phi}}_I\Phi_I-\bar{\Phi}_I\dot{\Phi}_I\right)=-2\nabla^\mu\Big{[}\bar{\Psi}_1\Psi_1m_1\cos^2\theta+\bar{\Psi}_2\Psi_2m_2\sin^2\theta\nonumber\\
&+\frac{1}{2}\left(\bar{\Psi}_1\Psi_2+\bar{\Psi}_2\Psi_1\right)(m_1+m_2)\cos\theta\sin\theta\Big{]}_{\mathcal{O}(\hbar)}+\mathcal{O}(\hbar^2),
\end{align}
\begin{align}\label{AppEIntermediate2}
&i\hbar\left(\nabla^\mu{\bar\Phi}_I\dot{\Phi}_I-\dot{\bar{\Phi}}_I\nabla^\mu\Phi_I\right)=\left[-(\nabla_\nu \varv^\nu)\bar{\Phi}_I{\Phi}_I\,p_I^\mu\right.\nonumber\\
&\left.+m_1\nabla^\mu\left(\bar{\Psi}_1\Psi_1\right)\cos^2\theta+m_2\nabla^\mu\left(\bar{\Psi}_2\Psi_2\right)\sin^2\theta\right]_{\mathcal{O}(\hbar)}\nonumber\\
&+\left(m_2\nabla^\mu\bar{\Psi}_1\Psi_2+m_1\bar{\Psi}_1\nabla^\mu\Psi_2+m_2\bar{\Psi}_2\nabla^\mu\Psi_1+m_1\nabla^\mu\bar{\Psi}_2\Psi_1\right)_{\mathcal{O}(\hbar)}\nonumber\\
&\quad\times\cos\theta\sin\theta+\mathcal{O}(\hbar^2).
\end{align}
For convenience, we also introduced here the notation $\mathcal{S}_{ij}=\mathcal{S}_i-\mathcal{S}_j$. 
Inserting these latter results, together with Eq.\,(\ref{DotPhiPhi}), into Eq.\,(\ref{AppEFirstStepDotP}), we find, up to first order in $\hbar$:
\begin{align}\label{AppCFinalDotPe}
    \dot{p}_I^{A\mu}&=-\frac{1}{2} R^{\mu}_{\;\;\nu\rho\sigma}\varv^\nu S_I^{\rho\sigma}-(\nabla^\mu \varv_{\nu})p_I^{A\nu}\nonumber\\
    &\quad+\Delta m_{21}\sin2\theta\left(\frac{i}{\hbar}\frac{\bar{\Psi}_1\Psi_2-\bar{\Psi}_2\Psi_1}{2\bar{\Phi}_I\Phi_I}\, p_I^{A\mu}\right.\nonumber\\
&\quad\left.+\frac{\nabla^\mu\bar{\Psi}_1\Psi_2-\bar{\Psi}_1\nabla^\mu\Psi_2+\bar{\Psi}_2\nabla^\mu\Psi_1-\nabla^\mu\bar{\Psi}_2\Psi_1}{4\bar{\Phi}_I\Phi_I}\right)_{\mathcal{O}(\hbar)}\!\!\!\!+\mathcal{O}(\hbar^2).
\end{align}

\section{Derivation of Eq.\,(\ref{DotS})}\label{App6}
The strategy for proving Eq.\,(\ref{DotS}) is to also work first with $\Phi$ before switching to its components $\Psi_1$ and $\Psi_2$ and the spinors $\psi_i^{(0)}(x)$ and $\psi_i^{(1)}(x)$ in their respective WKB expansions. Using the definition (\ref{SpinS}) of $S^{\mu\nu}$, the expansions (\ref{AppEPhi&DotPhi}), as well as Eq.\,(\ref{DotPhiPhi}), we find, after keeping only terms of the order $\hbar$, that
\begin{align}\label{AppBFirstWorkWithPsie}
    \dot{ S}_I^{\mu\nu}&=\hbar\frac{\dot{\bar{\Phi}}_I\sigma^{\mu\nu}\Phi_I+\bar{\Phi}_I\sigma^{\mu\nu}\dot{\Phi}_I}{2\bar{\Phi}_I\Phi_I}-\frac{\overset{\boldsymbol .}{(\bar{\Phi}_I\Phi_I)}}{\bar{\Phi}_I{\Phi}_I}S_I^{\mu\nu}\nonumber\\
&=-(\nabla_\rho \varv^\rho)S_I^{\mu\nu}+i\left(\frac{\bar{\Psi}_2\sigma^{\mu\nu}\Psi_1-\bar{\Psi}_1\sigma^{\mu\nu}\Psi_2}{4\bar{\Phi}_I\Phi_I}\right)_{\mathcal{O}(\hbar)}\Delta m_{21}\sin2\theta\nonumber\\
&\quad+\left[\nabla_\rho \varv^\rho+\left(\frac{i}{\hbar}\frac{\bar{\Psi}_1\Psi_2-\bar{\Psi}_2\Psi_1}{2\bar{\Phi}_I\Phi_I}\right)_{\mathcal{O}(\hbar)}\Delta m_{21}\sin2\theta\right]S_I^{\mu\nu}\nonumber\\
&=\left[i\left(\frac{\bar{\Psi}_2\sigma^{\mu\nu}\Psi_1-\bar{\Psi}_1\sigma^{\mu\nu}\Psi_2}{4\bar{\Phi}_I\Phi_I}\right)_{\mathcal{O}(\hbar)}\right.\nonumber\\
&\qquad\left.+\left(\frac{i}{\hbar}\frac{\bar{\Psi}_1\Psi_2-\bar{\Psi}_2\Psi_1}{2\bar{\Phi}_I\Phi_I}\right)_{\mathcal{O}(\hbar)}\, S_I^{\mu\nu}\right]\Delta m_{21}\sin2\theta+\mathcal{O}(\hbar^2).
\end{align}
\section{Derivation of Eq.\,(\ref{DotPOption2})}\label{App7}
We give in this appendix the detailed steps leading to Eq.\,(\ref{DotPOption2}). Starting from the definition (\ref{pIOption2}) of the dynamical 4-momentum $p_I^{B\mu}(x)$, carried by the spinor field $\Phi_I(x)$, we compute the proper time derivative $\dot{p}_I^{B\mu}(x)$ by performing the same terms rearrangements we performed to extract Eqs.\,(\ref{AppBFirstStep}) and (\ref{AppBSecondStep}). The result is:
\begin{align}\label{App7FirstStepDotP}
\dot{p}_I^{B\mu}&=\frac{i\hbar m_I m_{II}}{m_Im_{II}-m^2_{I,II}}\frac{\rm d}{{\rm d}\tau}\left(\frac{\nabla^\mu\bar{\Phi}_I\Phi_I-\bar{\Phi}_I\nabla^\mu\Phi_I}{2\bar{\Phi}_I\Phi_I}\right.\nonumber\\
&\qquad\qquad\qquad\qquad\qquad\left.-\frac{m_{I,II}}{m_{II}}\frac{\nabla^\mu\bar{\Phi}_{II}\Phi_{I}-\bar{\Phi}_{I}\nabla^\mu\Phi_{II}}{2\bar{\Phi}_I\Phi_I}\right)\nonumber\\
    &=-\frac{\overset{\boldsymbol .}{(\bar{\Phi}_I\Phi_I)}}{\bar{\Phi}_I\Phi_I}p_I^{B\mu}-(\nabla^\mu \varv_{\nu})p_I^{B\nu}\nonumber\\
    &-\frac{m_I m_{II}R^{\mu}_{\;\;\nu\rho\lambda}\varv^\nu}{2(m_Im_{II}-m^2_{I,II})}\left(S_I^{\rho\lambda}-\frac{\hbar m_{I,II}}{m_{II}}\frac{\bar{\Phi}_{II}\sigma^{\rho\lambda}\Phi_I+\bar{\Phi}_I\sigma^{\rho\lambda}\Phi_{II}}{4\bar{\Phi}_I\Phi_I}\right)\nonumber\\
    &+\frac{i\hbar m_I m_{II}}{m_Im_{II}-m^2_{I,II}}\left[ \frac{\nabla^\mu(\dot{\bar{\Phi}}_I\Phi_I-\bar{\Phi}_I\dot{\Phi}_I)}{2\bar{\Phi}_I\Phi_I}+ \frac{\nabla^\mu\bar{\Phi}_I\dot{\Phi}_I-\dot{\bar{\Phi}}_I\nabla^\mu\Phi_I}{\bar{\Phi}_I\Phi_I}\right]\nonumber\\
    &-\frac{i\hbar m_I m_{I,II}}{m_Im_{II}-m^2_{I,II}}\left[ \frac{\nabla^\mu(\dot{\bar{\Phi}}_{II}\Phi_I-\bar{\Phi}_I\dot{\Phi}_{II})}{2\bar{\Phi}_I\Phi_I}\right.\nonumber\\
    &\left.+ \frac{\nabla^\mu\bar{\Phi}_{II}\dot{\Phi}_I-\dot{\bar{\Phi}}_I\nabla^\mu\Phi_{II}+\nabla^\mu\bar{\Phi}_I\dot{\Phi}_{II}-\dot{\bar{\Phi}}_{II}\nabla^\mu\Phi_I}{2\bar{\Phi}_I\Phi_I}\right].
\end{align}
Using expression (\ref{PhiI&PhiII}) of the spinor field $\Phi_{II}(x)$, we have:
\begin{align}\label{App7Phi&DotPhi}
\Phi_{II}&=e^{\frac{i}{\hbar}\mathcal{S}_2}\left(\psi_2^{(0)}+\hbar\psi_2^{(1)}+\hbar^2\psi_2^{(2)}+\dots\right)\cos\theta\nonumber\\
&\quad-e^{\frac{i}{\hbar}\mathcal{S}_1}\left(\psi_1^{(0)}+\hbar\psi_1^{(1)}+\hbar^2\psi_1^{(2)}+\ldots\right)\sin\theta,\nonumber\\
\dot{\Phi}_{II}&=e^{\frac{i}{\hbar}\mathcal{S}_2}\left[\dot{\psi}_2^{(0)}+\hbar\dot{\psi}_2^{(1)}+\hbar^2\dot{\psi}_2^{(2)}+\ldots\right.\nonumber\\
&\quad\left.-\tfrac{i}{\hbar}m_2(\psi_2^{(0)}+\hbar\psi_2^{(1)}+\hbar^2\psi_2^{(2)}+\ldots)\right]\cos\theta\nonumber\\
&\quad-e^{\frac{i}{\hbar}\mathcal{S}_1}\left[\dot{\psi}_1^{(0)}+\hbar\dot{\psi}_1^{(1)}+\hbar^2\dot{\psi}_1^{(2)}+\ldots\right.\nonumber\\
&\quad\left.-\tfrac{i}{\hbar}m_1\left(\psi_1^{(0)}+\hbar\psi_1^{(1)}+\hbar^2\psi_1^{(2)}+\ldots\right)\right]\sin\theta.
\end{align}
In a similar manner to the calculation leading to Eq.\,(\ref{AppEIntermediate1}), we compute the numerators in the third term and in the last line on the right-hand side of Eq.\,(\ref{App7FirstStepDotP}), up to first order in $\hbar$, to be, respectively:
\begin{align}\label{AppGIntermediate1}
&i\hbar\nabla^\mu\left(\dot{\bar{\Phi}}_{II}\Phi_I-\bar{{\Phi}}_{I}\dot{\Phi}_{II}\right)=i\hbar(-\tfrac{1}{2}\nabla_\nu \varv^\nu)\nabla^\mu\Big{(}\bar{\Psi}_2\Psi_1-\bar{\Psi}_1\Psi_2\Big{)}\nonumber\\
&\qquad\qquad-(m_2\cos^2\theta-m_1\sin^2\theta)\nabla^\mu\left(\bar{\Psi}_2\Psi_1+\bar{\Psi}_1\Psi_2\right)\nonumber\\
&\qquad\qquad-2\nabla^\mu\left(m_2\bar{\Psi}_2\Psi_2-m_1\bar{\Psi}_1\Psi_1\right)\cos\theta\sin\theta,
\end{align}
\begin{align}\label{AppGIntermediate2}
&i\hbar\left(\nabla^\mu\bar{\Phi}_I\dot{\Phi}_{II}-\dot{\bar{\Phi}}_{II}\nabla^\mu\Phi_I+\nabla^\mu\bar{\Phi}_{II}\dot{\Phi}_I-\dot{\bar{\Phi}}_I\nabla^\mu\Phi_{II}\right)\nonumber\\[7pt]
&=-i\hbar(\tfrac{1}{2}\nabla_\nu \varv^\nu)\left(\nabla^\mu\bar{\Psi}_1\Psi_2-\bar{\Psi}_1\nabla^\mu\Psi_2+\nabla^\mu\bar{\Psi}_2\Psi_1-\bar{\Psi}_2\nabla^\mu\Psi_1\right)\nonumber\\
&\quad\times(\cos^2\theta-\sin^2\theta)-2\nabla^\mu\left(m_1\bar{\Psi}_1\Psi_1-m_2\bar{\Psi}_2\Psi_2\right)\cos\theta\sin\theta\nonumber\\
&\quad+i\hbar(\nabla_\nu \varv^\nu)(\bar{\Psi}_2\nabla^\mu\Psi_2-\nabla^\mu\bar{\Psi}_2\Psi_2+\nabla^\mu\bar{\Psi}_1\Psi_1-\bar{\Psi}_1\nabla^\mu\Psi_1)\nonumber\\
&\quad\times\cos\theta\sin\theta+\left[m_1\left(\nabla^\mu\bar{\Psi}_2\Psi_1+\bar{\Psi}_1\nabla^\mu\Psi_2\right)\right.\nonumber\\
&\quad\left.+m_2\left(\bar{\Psi}_2\nabla^\mu\Psi_1+\nabla^\mu\bar{\Psi}_1\Psi_2\right)\right](\cos^2\theta-\sin^2\theta).
\end{align}
Combining these last two identities with the results (\ref{DotPhiPhi}), (\ref{AppEIntermediate1}) and (\ref{AppEIntermediate2}), and inserting these all into Eq.\,(\ref{App7FirstStepDotP}), we find
\begin{align}\label{App7FinalDotP}
\dot{p}_I^{B\mu}&=-\frac{m_I m_{II}R^{\mu}_{\;\;\nu\rho\lambda}\varv^\nu}{2(m_Im_{II}\!-\!m^2_{I,II})}\!\left(\!S_I^{\rho\lambda}\!-\!\frac{\hbar m_{I,II}}{m_{II}}\frac{\bar{\Phi}_{II}\sigma^{\rho\lambda}\Phi_I\!+\!\bar{\Phi}_I\sigma^{\rho\lambda}\Phi_{II}}{4\bar{\Phi}_I\Phi_I}\!\right)_{\mathcal{O}(\hbar)}\nonumber\\
&\quad-(\nabla^\mu \varv_{\nu})p_I^{B\nu}\nonumber\\
    &+\!\left[\frac{m_I m_{II}}{m_Im_{II}\!-\!m^2_{I,II}}\frac{\nabla^\mu\bar{\Psi}_1\Psi_2\!-\!\bar{\Psi}_1\nabla^\mu\Psi_2\!+\!\bar{\Psi}_2\nabla^\mu\Psi_1\!-\!\nabla^\mu\bar{\Psi}_2\Psi_1}{4\bar{\Phi}_I\Phi_I}\right.\nonumber\\
    &\quad\left.+\frac{i}{\hbar}\frac{\bar{\Psi}_1\Psi_2-\bar{\Psi}_2\Psi_1}{2\bar{\Phi}_I\Phi_I}p_I^{B\mu}\right]_{\mathcal{O}(\hbar)}\Delta m_{21}\sin2\theta\nonumber\\
    &\quad+\frac{m_I m_{I,II}\Delta m_{21}}{m_Im_{II}-m^2_{I,II}}\left(\frac{\nabla^\mu\bar{\Psi}_2\Psi_1+\bar{\Psi}_1\nabla^\mu\Psi_2}{2\bar{\Phi}_I\Phi_I}\cos^2\theta\right.\nonumber\\    &\quad\left.+\frac{\nabla^\mu\bar{\Psi}_1\Psi_2+\bar{\Psi}_2\nabla^\mu\Psi_1}{2\bar{\Phi}_I\Phi_I}\sin^2\theta\right)_{\mathcal{O}(\hbar)}+\mathcal{O}(\hbar^2).
\end{align}

\section{Derivation of Eq.\,(\ref{DotPOption3})}\label{App8}
We give in this appendix the detailed steps leading to Eq.\,(\ref{DotPOption3}). Starting from the definition (\ref{POption3}) of the dynamical 4-momentum $p^\mu(x)$, carried by the superposition $\Sigma$, we compute the proper time derivative $\dot{p}^\mu(x)$ by performing the same terms rearrangements we performed to extract Eqs.\,(\ref{AppBFirstStep}) and (\ref{AppBSecondStep}). The result is:
\begin{align}\label{App8FirstStepDotPOption3}
&\dot{p}_{I,II}^\mu=\frac{m_I m_{II}}{\sqrt{m_1m_2}}\frac{\rm d}{{\rm d}\tau}\left(\frac{i\hbar}{m_I}\frac{\nabla^\mu\bar{\Phi}_I\Phi_I-\bar{\Phi}_I\nabla^\mu\Phi_I}{2\bar{\Sigma}\Sigma}\right.\nonumber\\
&\qquad\left.+\frac{i\hbar}{m_{II}}\frac{\nabla^\mu\bar{\Phi}_{II}\Phi_{II}-\bar{\Phi}_{II}\nabla^\mu\Phi_{II}}{2\bar{\Sigma}\Sigma}\right)\nonumber\\
 &-\frac{i\hbar m_{I,II}}{\sqrt{m_1m_2}}\frac{\rm d}{{\rm d}\tau}\left(\frac{\nabla^\mu\bar{\Phi}_{II}\Phi_{I}-\bar{\Phi}_{I}\nabla^\mu\Phi_{II}+\nabla^\mu\bar{\Phi}_{I}\Phi_{II}-\bar{\Phi}_{II}\nabla^\mu\Phi_{I}}{2\bar{\Sigma}\Sigma}\right)\nonumber\\ 
 &=-\frac{\overset{\boldsymbol .}{(\bar{\Sigma}\Sigma)}}{\bar{\Sigma}\Sigma}p_{I,II}^\mu-(\nabla^\mu \varv_\nu)p_{I,II}^\nu\nonumber\\
 &\quad-\tfrac{1}{2} R^{\mu}_{\;\;\nu\rho\lambda}\varv^\nu\left(\frac{m_{II}\bar{\Phi}_{I}\Phi_{I}}{\sqrt{m_1m_2}\bar{\Sigma}\Sigma}S_{I}^{\rho\lambda}+\frac{m_{I}\bar{\Phi}_{II}\Phi_{II}}{\sqrt{m_1m_2}\bar{\Sigma}\Sigma}S_{II}^{\rho\lambda}\right.\nonumber\\
 &\qquad\left.-\hbar m_{I,II}\frac{\bar{\Phi}_{II}\sigma^{\rho\lambda}\Phi_{I}+\bar{\Phi}_{I}\sigma^{\rho\lambda}\Phi_{II}}{2\sqrt{m_1m_2}\bar{\Sigma}\Sigma}\right)\nonumber\\
&\quad+\frac{i\hbar m_{II}}{\sqrt{m_1m_2}} \left[ \frac{\nabla^\mu(\dot{\bar{\Phi}}_I\Phi_I-\bar{\Phi}_I\dot{\Phi}_I)}{2\bar{\Sigma}\Sigma}+ \frac{\nabla^\mu\bar{\Phi}_I\dot{\Phi}_I-\dot{\bar{\Phi}}_I\nabla^\mu\Phi_I}{\bar{\Sigma}\Sigma}\right]\nonumber\\
&\quad+\frac{i\hbar m_{I}}{\sqrt{m_1m_2}} \left[ \frac{\nabla^\mu(\dot{\bar{\Phi}}_{II}\Phi_{II}-\bar{\Phi}_{II}\dot{\Phi}_{II})}{2\bar{\Sigma}\Sigma}+ \frac{\nabla^\mu\bar{\Phi}_{II}\dot{\Phi}_{II}-\dot{\bar{\Phi}}_{II}\nabla^\mu\Phi_{II}}{\bar{\Sigma}\Sigma}\right]\nonumber\\
&\quad-\frac{i\hbar m_{I,II}}{\sqrt{m_1m_2}}\left[ \frac{\nabla^\mu(\dot{\bar{\Phi}}_{II}\Phi_I-\bar{\Phi}_I\dot{\Phi}_{II}+\dot{\bar{\Phi}}_I\Phi_{II}-\bar{\Phi}_{II}\dot{\Phi}_{I})}{2\bar{\Sigma}\Sigma}\right]\nonumber\\
&\quad-\frac{i\hbar m_{I,II}}{\sqrt{m_1m_2}}\left( \frac{\nabla^\mu\bar{\Phi}_{II}\dot{\Phi}_I-\dot{\bar{\Phi}}_I\nabla^\mu\Phi_{II}+\nabla^\mu\bar{\Phi}_I\dot{\Phi}_{II}-\dot{\bar{\Phi}}_{II}\nabla^\mu\Phi_I}{\bar{\Sigma}\Sigma}\right).
\end{align}
First, using the result (\ref{DotPhiPhi}) together with the time derivative $\frac{\rm d}{\rm d\tau}(\bar{\Phi}_{II}\Phi_{II})$, obtained from Eq.\,(\ref{DotPhiPhi}) by the substitutions $I\rightarrow II$, $\theta\rightarrow -\theta$ and $1\leftrightarrow2$, we learn that $\frac{\rm d}{\rm d\tau}(\bar{\Sigma}\Sigma)=\frac{\rm d}{\rm d\tau}(\bar{\Phi}_I\Phi_I+\bar{\Phi}_{II}\Phi_{II})=-(\nabla_\mu \varv^\mu)(\bar{\Sigma}\Sigma)_{\mathcal{O}(\hbar)}+\mathcal{O}(\hbar^2)$. With this latter identity, together with identities (\ref{AppEIntermediate1}), (\ref{AppEIntermediate2}), (\ref{AppGIntermediate1}) and (\ref{AppGIntermediate2}), Eq.\,(\ref{App8FirstStepDotPOption3}) reduces to
\begin{align}\label{App8DotPOption3}
\dot{p}_{I,II}^{\mu}&=-\tfrac{1}{2} R^{\mu}_{\;\;\nu\rho\lambda}\varv^\nu\left(\frac{m_{II}\bar{\Phi}_{I}\Phi_{I}}{\sqrt{m_1m_2}\bar{\Sigma}\Sigma}S_{I}^{\rho\lambda}+\frac{m_{I}\bar{\Phi}_{II}\Phi_{II}}{\sqrt{m_1m_2}\bar{\Sigma}\Sigma}S_{II}^{\rho\lambda}\right.\nonumber\\
&\quad\left.-\hbar m_{I,II}\frac{\bar{\Phi}_{II}\sigma^{\rho\lambda}\Phi_{I}+\bar{\Phi}_{I}\sigma^{\rho\lambda}\Phi_{II}}{2\sqrt{m_1m_2}\bar{\Sigma}\Sigma}\right)_{\mathcal{O}(\hbar)}-(\nabla^\mu \varv_\nu)p_{I,II}^\nu\nonumber\\
&\quad+\mathcal{O}(\hbar^2).
\end{align}
{\color{white}....}\\
{\color{white}....}\\
{\color{white}....}\\
{\color{white}....}\\
{\color{white}....}\\
{\color{white}....}\\
{\color{white}....}\\
{\color{white}....}\\
{\color{white}....}\\
{\color{white}....}\\
{\color{white}....}\\
{\color{white}....}\\
{\color{white}....}\\
{\color{white}....}\\
{\color{white}....}\\
{\color{white}....}\\
{\color{white}....}

\end{document}